\documentclass[sn-nature]{sn-jnl}

\usepackage{hyperref}
\usepackage{graphicx}%
\usepackage{caption}
\usepackage{multirow}%
\usepackage{amsmath,amssymb,amsfonts}%
\usepackage{amsthm}%
\usepackage{mathrsfs}%
\usepackage[title]{appendix}%
\usepackage{xcolor}%
\usepackage{textcomp}%
\usepackage{manyfoot}%
\usepackage{booktabs}%
\usepackage{algorithm}%
\usepackage{algorithmicx}%
\usepackage{algpseudocode}%
\usepackage{listings}%
\usepackage{longtable}
\usepackage{gensymb}
\usepackage{pdflscape}
\usepackage{verbatim}
\usepackage{anyfontsize}
\usepackage[utf8]{inputenc}

\theoremstyle{thmstyleone}%

\theoremstyle{thmstyletwo}%

\theoremstyle{thmstylethree}%

\newcommand{\quotes}[1]{``#1''}

\begin{document}

\title[Catastrophic disruption of asteroid 2023 CX1]{Catastrophic disruption of asteroid 2023 CX1 and implications for planetary defense}

\author*[1,2,3]{\fnm{Auriane} \sur{Egal}}
\email{auriane.egal@obspm.fr, auriane.egal@montreal.ca}
\author[3]{\fnm{Denis} \sur{Vida}}
\author[1]{\fnm{Fran\c{c}ois} \sur{Colas}}
\author[1,4]{\fnm{Brigitte} \sur{Zanda}}
\author[1,5]{\fnm{Sylvain} \sur{Bouley}}
\author[1,6]{\fnm{Asma} \sur{Steinhausser}}
\author[1,7]{\fnm{Pierre} \sur{Vernazza}}
\author[8,9]{\fnm{Ludovic} \sur{Ferri\`{e}re}}
\author[10]{\fnm{J\'{e}r\^{o}me} \sur{Gattacceca}}
\author[1,11]{\fnm{Mirel} \sur{Birlan}}
\author[1]{\fnm{J\'{e}r\'{e}mie} \sur{Vaubaillon}}
\author[12]{\fnm{Karl} \sur{Antier}}
\author[1,11]{\fnm{Simon} \sur{Anghel}}
\author[1,13]{\fnm{Josselin} \sur{Desmars}}
\author[1]{\fnm{K\'{e}vin} \sur{Bailli\'{e}}}
\author[1]{\fnm{Lucie} \sur{Maquet}}
\author[14]{\fnm{S\'{e}bastien} \sur{Bouquillon}}
\author[15]{\fnm{Adrien} \sur{Malgoyre}}
\author[1]{\fnm{Simon} \sur{Jeanne}}
\author[]{\fnm{FRIPON} \sur{International Team\textsuperscript{\textdagger}}}
\author[16]{\fnm{Ji\v{r}\'i} \sur{Borovi\v{c}ka}}
\author[16]{\fnm{Pavel} \sur{Spurn\'y}}
\author[17]{\fnm{Hadrien A. R.} \sur{Devillepoix}}
\author[18]{\fnm{Marco} \sur{Micheli}}
\author[19]{\fnm{Davide} \sur{Farnocchia}}
\author[19]{\fnm{Shantanu} \sur{Naidu}}
\author[3]{\fnm{Peter} \sur{Brown}}
\author[3]{\fnm{Paul} \sur{Wiegert}}
\author[20]{\fnm{Kriszti\'an} \sur{S\'arneczky}}
\author[21]{\fnm{Andr\'as} \sur{P\'al}}
\author[22]{\fnm{Nick} \sur{Moskovitz$^\text{22}$}}
\author[22]{\fnm{Theodore} \sur{Kareta$^\text{22}$}}
\author[23,24]{\fnm{Toni} \sur{Santana-Ros}}
\author[25]{\fnm{Alexis} \sur{Le Pichon}}
\author[25]{\fnm{Gilles} \sur{Mazet-Roux}}
\author[25]{\fnm{Julien} \sur{Vergoz}}
\author[3]{\fnm{Luke} \sur{McFadden}}
\author[26]{\fnm{Jelle} \sur{Assink}}
\author[26,27]{\fnm{Läslo} \sur{Evers}}
\author[28]{\fnm{Daniela} \sur{Krietsch}}
\author[28]{\fnm{Henner} \sur{Busemann}}
\author[28]{\fnm{Colin} \sur{Maden}}
\author[28]{\fnm{Lisa Maria} \sur{Eckart}}
\author[29,30]{\fnm{Jean-Alix} \sur{Barrat}}
\author[31]{\fnm{Pavel} \sur{Povinec}}
\author[31]{\fnm{Ivan} \sur{S\'ykora}}
\author[31]{\fnm{Ivan} \sur{Kontul'}}
\author[32]{\fnm{Oscar} \sur{Marchhart}}
\author[32]{\fnm{Martin} \sur{Martschini}}
\author[32]{\fnm{Silke} \sur{Merchel}}
\author[32]{\fnm{Alexander} \sur{Wieser}}
\author[4]{\fnm{Matthieu} \sur{Gounelle}}
\author[4]{\fnm{Sylvain} \sur{Pont}}
\author[4]{\fnm{Pierre} \sur{Sans-Jofre}}
\author[33]{\fnm{Sebastiaan} \sur{de Vet$^\text{33,34}$}}
\author[35]{\fnm{Ioannis} \sur{Baziotis}}
\author[36]{\fnm{Miroslav} \sur{Brož}}
\author[37]{\fnm{Michaël} \sur{Marsset}}
\author[38]{\fnm{J\'{e}r\^{o}me} \sur{Vergne}}
\author[36]{\fnm{Josef} \sur{Hanu\v{s}}}
\author[18]{\fnm{Maxime} \sur{Devog\`{e}le}}
\author[18]{\fnm{Luca} \sur{Conversi}}
\author[18]{\fnm{Francisco} \sur{Ocaña}}
\author[39]{\fnm{Luca} \sur{Buzzi}}
\author[11]{\fnm{Dan Alin} \sur{Nedelcu$^\text{11}$}}
\author[11]{\fnm{Adrian} \sur{Sonka$^\text{11}$}}
\author[40]{\fnm{Florent} \sur{Losse}}
\author[41]{\fnm{Philippe} \sur{Dupouy}}
\author[42]{\fnm{Korado} \sur{Korlevi\'{c}}}
\author[43]{\fnm{Dieter} \sur{Husar}}
\author[43]{\fnm{Jost} \sur{Jahn}}
\author[44]{\fnm{Damir} \sur{\v{S}egon$^\text{44}$}}
\author[45]{\fnm{Mark} \sur{McIntyre}}
\author[46]{\fnm{Ralf} \sur{Neubert}}
\author[47]{\fnm{Pierre} \sur{Beck}}
\author[1,17]{\fnm{Patrick} \sur{Shober}}
\author[10]{\fnm{Anthony} \sur{Lagain}}
\author[2]{\fnm{Olivier} \sur{Hernandez}}
\author[48]{\fnm{Darrel} \sur{Robertson}}
\author[49]{\fnm{Peter} \sur{Jenniskens}}

\affil[1]{\orgdiv{FRIPON}, \orgname{LTE, Observatoire de Paris, PSL Research University, Sorbonne Universit\'{e}, CNRS, Universit\'{e} Lille}, \orgaddress{\street{c/o 77 av. Denfert-Rochereau},  \city{Paris}, \postcode{75014}, \country{France}}}
\affil[2]{\orgdiv{Plan\'etarium de Montr\'eal}, \orgname{Espace pour la Vie}, \orgaddress{\street{4801 av. Pierre-de Coubertin},  \city{Montr\'eal}, \postcode{H1V 3V4}, \state{Qu\'ebec}, \country{Canada}}}
\affil[3]{\orgdiv{Department of Physics and Astronomy}, \orgname{The University of Western Ontario}, \orgaddress{\street{1151 Richmond Street},  \city{London}, \postcode{N6A 3K7}, \state{Ontario}, \country{Canada}}}
\affil[4]{\orgdiv{Mus\'{e}um National d'Histoire Naturelle}, \orgname{IMPMC}, \orgaddress{\street{61 rue Buffon},  \city{Paris}, \postcode{75005}, \country{France}}}
\affil[5]{\orgdiv{GEOPS}, \orgname{Universit\'{e} Paris-Saclay}, \orgaddress{\street{Rue du Belv\'{e}d\`{e}re B\^{a}timent 504},  \city{Orsay}, \postcode{91400}, \country{France}}}
\affil[6]{\orgdiv{PatriNat}, \orgname{MNHN, OFB, IRD, CNRS}, \orgaddress{\street{36 rue Geoffroy-Saint-Hilaire},  \city{Paris}, \postcode{75005}, \country{France}}}
\affil[7]{\orgname{Laboratoire d'Astrophysique de Marseille}, \orgaddress{\street{38 Rue Fr\'{e}d\'{e}ric Joliot Curie},  \city{Marseille}, \postcode{13013}, \country{France}}}
\affil[8]{\orgname{Naturhistorisches Museum Wien}, \orgaddress{\street{Burgring 7},  \city{Vienna}, \postcode{1010}, \country{Austria}}}
\affil[9]{\orgname{Natural History Museum Abu Dhabi}, \orgaddress{\street{Jacques Chirac street,  Saadiyat Island},  \city{Abu Dhabi}, \country{United Arab Emirates}}}
\affil[10]{\orgname{CNRS, Aix-Marseille Univ, IRD, INRAE, CEREGE}, \orgaddress{ \city{Aix-en-Provence}, \country{France}}}
\affil[11]{\orgname{Astronomical Institute of the Romanian Academy}, \orgaddress{\street{5 Cu\c{t}itul de Argint},  \city{Bucharest}, \postcode{40557}, \country{Romania}}}
\affil[12]{\orgname{International Meteor Organization}, \orgaddress{\street{Mattheessensstraat 60},  \city{Hove}, \postcode{2540}, \country{Belgium}}}
\affil[13]{\orgname{Institut Polytechnique des Sciences Avanc\'{e}es IPSA}, \orgaddress{\street{63 boulevard de Brandebourg},  \city{Ivry-sur-Seine}, \postcode{F-94200}, \country{France}}}
\affil[14]{\orgdiv{SYRTE, CNRS}, \orgname{Observatoire de Paris, Universit\'{e} PSL, Sorbonne Universit\'{e}s}, \orgaddress{\street{61 avenue de l’Observatoire},  \city{Paris}, \postcode{75014}, \country{France}}}
\affil[15]{\orgdiv{Institut Pytheas}, \orgname{Observatoire des Sciences de l'Univers}, \orgaddress{\street{38 Rue Fr\'{e}d\'{e}ric Joliot Curie},  \city{Marseille}, \postcode{13013}, \country{France}}}
\affil[16]{\orgname{Astronomical Institute of the Czech Academy of Sciences}, \orgaddress{\street{Fri\v{c}ova 298},  \city{Ond\v{r}ejov}, \postcode{25165}, \country{Czech Republic}}}
\affil[17]{\orgdiv{Space Science and Technology Centre and International Centre for Radio Astronomy Research}, \orgname{Curtin University}, \orgaddress{\street{GPO Box U1987},  \city{Perth}, \postcode{6845}, \state{WA}, \country{Australia}}}
\affil[18]{\orgdiv{ESA NEO Coordination Centre}, \orgname{Planetary Defence Office}, \orgaddress{\street{Largo Galileo Galilei, 1},  \city{Frascati (RM)}, \postcode{00044}, \country{Italy}}}
\affil[19]{\orgname{Jet Propulsion Laboratory, California Institute of Technology}, \orgaddress{\street{4800 Oak Grove Dr.},  \city{Pasadena}, \postcode{91109}, \state{CA}, \country{USA}}}
\affil[20]{\orgdiv{18, Konkoly Observatory}, \orgname{HUN-REN Research Centre for Astronomy and Earth Sciences}, \orgaddress{\street{Konkoly-Thege Miklós út 15-17},  \city{Budapest}, \postcode{H-1121}, \country{Hungary}}}
\affil[21]{\orgdiv{Eötvös Loránd University}, \orgname{Institute of Physics and Astronomy}, \orgaddress{\street{H-1117},  \city{Budapest}, \postcode{38000}, \country{Hungary}}}
\affil[22]{\orgname{Lowell Observatory}, \orgaddress{\street{1400 W Mars Hill Road},  \city{Flagstaff}, \postcode{86001}, \state{AZ}, \country{USA}}}
\affil[23]{\orgdiv{Departamento de Física, Ingeniería de Sistemas y Teoría de la Señal}, \orgname{Universidad de Alicante}, \orgaddress{\street{Carr. de San Vicente del Raspeig, s/n,},  \city{Alicante}, \postcode{03690}, \country{Spain}}}
\affil[24]{\orgdiv{Institut de Ci\`{e}ncies del Cosmos (ICCUB)}, \orgname{Universitat de Barcelona (IEEC-UB)}, \orgaddress{\street{Carrer de Mart\'{\i} i Franqu\`{e}s, 1},  \city{Barcelona}, \postcode{08028}, \country{Spain}}}
\affil[25]{\orgname{CEA}, \orgaddress{\street{DAM DIF},  \city{Arpajon}, \postcode{F-91297}, \country{France}}}
\affil[26]{\orgname{Royal Netherlands Meteorological Institute}, \orgaddress{\street{Utrechtseweg 297},  \city{De Bilt}, \postcode{3731 GA}, \country{The Netherlands}}}
\affil[27]{\orgname{Delft University of Technology}, \orgaddress{\street{Kluyverweg 1},  \city{Delft}, \postcode{2629HS}, \country{The Netherlands}}}
\affil[28]{\orgdiv{Inst. für Geochemie und Petrologie}, \orgname{ETH Zürich}, \orgaddress{\street{Clausiusstrasse 25},  \city{8092 Zürich} \country{Switzerland}}}
\affil[29]{\orgname{Univ. Brest, CNRS, Ifremer, IRD, LEMAR, Institut Universitaire Europ\'{e}en de la Mer (IUEM)}, \orgaddress{\street{rue Dumont d’Urville},  \city{Plouzan\'{e}}, \postcode{29280}, \country{France}}}
\affil[30]{\orgdiv{Institut Universitaire de France}, \orgname{Universit\'{e} de Bretagne Occidentale}, \orgaddress{\street{3 Rue Matthieu Gallou},  \city{Brest}, \postcode{29200}, \country{France}}}
\affil[31]{\orgdiv{Faculty of Mathematics, Physics and Informatics}, \orgname{Comenius University}, \orgaddress{\street{Mlynska Dolina F-1},  \city{Bratislava}, \postcode{84248}, \country{Slovakia}}}
\affil[32]{\orgdiv{University of Vienna}, \orgname{Faculty of Physics, Isotope Physics}, \orgaddress{\street{Waehringer Str. 17},  \city{Vienna}, \postcode{A-1090}, \country{Austria}}}
\affil[33]{\orgdiv{Faculty of Aerospace Engineering}, \orgname{Delft University of Technology}, \orgaddress{\street{Kluyverweg 1},  \city{Delft}, \postcode{2629HS}, \country{The Netherlands}}}
\affil[34]{\orgname{Naturalis Biodiversity Centre}, \orgaddress{\street{Darwinweg 2},  \city{Leiden}, \postcode{2333 CR}, \country{The Netherlands}}}
\affil[35]{\orgname{Agricultural University of Athens}, \orgaddress{\street{Iera Odos 75},  \city{Athens}, \postcode{11855}, \country{Greece}}}
\affil[36]{\orgdiv{Charles University}, \orgname{Faculty of Mathematics and Physics, Institute of Astronomy}, \orgaddress{\street{V~Hole{\v s}ovi{\v c}k{\'a}ch 2},  \city{Prague}, \postcode{18000}, \country{Czech Republic}}}
\affil[37]{\orgname{European Southern Observatory}, \orgaddress{\street{Karl-Schwarzschild-Straße 2},  \city{Garching}, \postcode{85748}, \country{Germany}}}
\affil[38]{\orgname{Universit\'{e} de Strasbourg}, \orgaddress{\street{5 rue Ren\'{e} Descartes},  \city{Strasbourg}, \postcode{67084}, \country{France}}}
\affil[39]{\orgname{Schiaparelli Observatory}, \orgaddress{\street{Campo dei Fiori},  \city{Varese}, \postcode{21100}, \country{Italy}}}
\affil[40]{\orgname{Saint-Pardon-de-Conques Observatory}, \orgaddress{\street{Rue des Roberts},  \city{Saint-Pardon-de-Conques}, \postcode{33210}, \country{France}}}
\affil[41]{\orgname{Dax Observatory}, \orgaddress{\street{Rue Pascal Lafitte},  \city{Dax}, \postcode{40100}, \country{France}}}
\affil[42]{\orgname{Vi\v{s}njan Observatory}, \orgaddress{\street{Istarska 5},  \city{Visnjan}, \postcode{52463-HR}, \country{Croatia}}}
\affil[43]{\orgname{SATINO Remote Observatory}, \orgaddress{\street{1912 route de l'Observatoire},  \city{Saint-Michel-L'Observatoire}, \postcode{04870}, \country{France}}}
\affil[44]{\orgname{Astronomical Society Istra Pula}, \orgaddress{\street{Monte Zaro 2},  \city{Pula}, \postcode{52100}, \country{Croatia}}}
\affil[45]{\orgname{Flint House}, \orgaddress{\street{40 Medcroft Road},  \city{Tackley}, \postcode{OX5 3AH}, \state{Oxfordshire}, \country{UK}}}
\affil[46]{\orgname{Th\"uringer Landessternwarte Tautenburg}, \orgaddress{\street{Sternwarte 5},  \city{Tautenburg}, \postcode{D-07778}, \country{Germany}}}
\affil[47]{\orgdiv{CNRS, IPAG}, \orgname{Univ. Grenoble Alpes}, \orgaddress{\street{414 rue de la piscine},  \city{Grenoble}, \postcode{38000}, \country{France}}}
\affil[48]{\orgdiv{Asteroid Threat Assessment Project}, \orgname{NASA Ames Research Center}, \orgaddress{\street{Mail Stop 244-11},  \city{Moffett Field}, \postcode{94035}, \state{CA}, \country{USA}}}
\affil[49]{\orgname{SETI Institute}, \orgaddress{\street{339 Bernardo Ave},  \city{Mountain View}, \postcode{94043}, \state{CA}, \country{USA}}}

 \abstract{Mitigation of the threat from airbursting asteroids requires an understanding of the potential risk they pose for the ground. How asteroids release their kinetic energy in the atmosphere is not well understood due to the rarity of significant impacts. Ordinary chondrites, in particular L chondrites, represent a frequent type of Earth-impacting asteroids. Here, we present the first comprehensive, space-to-lab characterization of an L chondrite impact. Small asteroid 2023 CX1 was detected in space and predicted to impact over Normandy, France, on 13 February 2023.  Observations from multiple independent sensors and reduction techniques revealed an unusual but potentially high-risk fragmentation behavior.  The nearly spherical 650 $\pm$ 160 kg (72 $\pm$ 6 cm diameter) asteroid catastrophically fragmented around 28 km altitude, releasing 98\% of its total energy in a concentrated region of the atmosphere. The resulting shockwave was spherical, not cylindrical, and released more energy closer to the ground. This type of fragmentation increases the risk of significant damage at ground level. These results warrant consideration for a planetary defense strategy for cases where a $>$3-4 MPa dynamic pressure is expected, including planning for evacuation of areas beneath anticipated disruption locations.}


\maketitle

\section{Introduction}

Energetic airbursting asteroid impacts such as the 500 kt of TNT equivalent Chelyabinsk event in February 2013 \cite{Brown2013} are exceedingly rare, leaving significant uncertainty about the altitude of energy deposition by different asteroid types. Among the most dangerous deeply penetrating asteroids are the S- and Q-class, which deliver ordinary chondrite meteorites. These meteorites are classified into H, L, and LL groups based on petrographic characteristics and distinct parent bodies. L chondrites are the most commonly recovered from observed falls \cite{Greenwood2020}, making them one of the most frequently impacting asteroid types.

On 12 February 2023, a small asteroid approaching Earth was discovered at the Piszk\'{e}stet\H{o} station of the Konkoly Observatory, in Hungary \cite{Sarneczky2024}. Designated 2023 CX1, the asteroid was predicted to impact 6.7 hours later over Normandy, France. Prompt alerts from ESA and NASA were quickly shared by the International Meteor Organization, as well as by the Fireball Recovery and InterPlanetary Observation Network (FRIPON) and its citizen science counterpart Vigie-Ciel \cite{Colas2014,Colas2020}. The rapid mobilization of these networks enabled the first large-scale, targeted observations of a fireball. A comprehensive dataset of optical, radio, infrasound, and seismic measurements was collected during the meteoroid's atmospheric entry at 02:59 UTC on 13 February.

Thanks to rapid data analysis and the exceptional mobilization of the public through the FRIPON/Vigie-Ciel network, a coordinated ground search led to the recovery of a first 93 g meteorite just two days after the fall, near the village of Saint-Pierre-le-Viger in Normandy \cite{JenniskensColas2023,Zanda2023}. The meteorite was named after this village and is hereafter referred to as SPLV. It is classified as an L5–6 chondrite breccia based on mineralogical and petrographic features, including an olivine Fa content of 24.6–26.5, low-Ni sulfides ($<$0.5~wt\%), orthorhombic low-Ca pyroxene, limited olivine variation ($<$5\%), and a scarcity of chondrules in thin sections. Shock indicators such as undulatory extinction, weak mosaicism and planar fractures in olivine, and abundant shock veins support a shock stage of S3 \cite{Gattacceca2024}. Further mineralogical details are provided in \cite{Bischoff2023}.

Comprehensive investigations of asteroid impacts —combining pre-impact observations, atmospheric fragmentation analysis, and geochemical study of recovered meteorites— remain rare. To date, only eleven asteroids have been detected prior to impact. Among them, just four led to meteorite recovery, and detailed atmospheric observations are lacking in most cases. SPLV is the only L chondrite linked to a pre-impact detection. Here, we present a multi-disciplinary analysis of SPLV, combining telescopic measurements, fireball observations, precise meteorite recovery, and laboratory analyses to investigate how L chondrite asteroids fragment and deposit energy in Earth’s atmosphere.

\section{Results}

\subsection{Asteroid orbit, shape and size}

\subsubsection{Orbit}

Between February 12, 20:18 UTC, and February 13, 02:52 UTC, 22 astronomical observatories reported observations of 2023 CX1 to the Minor Planet Center (MPC). High-fidelity orbit determination based on precise asteroid tracking enabled the prediction of the impact time to within a few seconds. Targeted observations of the fireball allowed for a detailed comparison between the telescopic and fireball-derived orbits by examining their relative positions at the top of the atmosphere. Initially, a significant offset of only $\sim$90~m was observed between the predicted impact location and the first fireball detection at an altitude of 101.755 km. A thorough analysis of the measurements revealed that the primary source of error stemmed from the confusion between WGS84 and mean sea level elevations of the station coordinates reported by the observers to the MPC. After correcting the coordinates, the asteroid's position at the reference altitude, derived from both telescopic and fireball observations, differed by less than 30 meters, well within $1\sigma$ uncertainties (Table \ref{tab:orbit} and Figure \ref{fig:orbit_comp}). This strong agreement enabled a cross-calibration between telescopic and fireball data, reducing the final relative offset between the trajectories at the reference altitude to approximately 18 meters (Table \ref{tab:orbit}), and establishing SPLV as one of the most precise meteorite orbits ever measured.

\begin{table}[h]
\caption{Orbital parameters and atmospheric entry location of 2023 CX1, determined from telescopic asteroid observations prior to impact (JPL \#13) and from in-atmosphere meteor data. The combined solution (JPL \#14) incorporates telescopic data, anchored by the atmospheric entry point derived from meteor observations. The orbital elements are for an osculating epoch of 2022-12-14.0 TDB, i.e., 60 days before impact, before significant perturbation from Earth's gravity. The meteoroid's location at a fixed reference altitude of 101.755 km, corresponding to the first observed position of the fireball, is provided for each orbital solution.}\label{tab:orbit}%
\resizebox{\textwidth}{!}{\begin{tabular}{@{}llrrr@{}}
\toprule
 Symbol & Parameter  & Telescopic (JPL \#13) & In-atmosphere & Combined (JPL \#14)\\
\midrule
$e$         & Eccentricity                             &   0.434567 $\pm$ 1e-5 &   0.4369 $\pm$ 2e-3 &   0.4345664 $\pm$ 6e-6  \\
$a$         &  Semimajor axis (au)                     &   1.629319 $\pm$ 4e-5 &   1.6356 $\pm$ 6e-3 &   1.629318 $\pm$ 1e-5  \\
$q$         &  Perihelion distance (au)                &   0.9212712 $\pm$ 2e-6 &   0.92106 $\pm$ 5e-4 &   0.92127115 $\pm$ 8e-7  \\
$i$         & Inclination ($\degree$)                  &   3.41838 $\pm$ 1e-4 &   3.430 $\pm$ 3e-2 &   3.418380 $\pm$ 4e-5  \\
$\Omega$    & Longitude of ascending node ($\degree$)  & 323.8709587 $\pm$ 5e-6 & 323.87032 $\pm$ 9e-4 & 323.8709589 $\pm$ 2e-6  \\
$\omega$    & Argument of perihelion ($\degree$)       & 218.78816 $\pm$ 2e-4 & 218.779 $\pm$ 7e-2 & 218.788170 $\pm$ 7e-5  \\
$m$         & Mean anomaly  ($\degree$)                & 316.5723 $\pm$ 1e-3 & 316.84 $\pm$ 2e-1 & 316.57224 $\pm$ 6e-4  \\
\midrule
$\phi$ & Latitude ($\degree$N) & 49.921614 $\pm$ 1e-4  & 49.92182 $\pm$ 2e-4 & 49.921654 $\pm$ 6e-5  \\
$\lambda$ & Longitude ($\degree$E) & -0.167195 $\pm$ 4e-4   & -0.16713 $\pm$ 2e-4 & -0.167172 $\pm$ 1e-4    \\
H & Altitude (km) & 101.755 $\pm$ 0.00 & 101.755 $\pm$ 0.00 & 101.755 $\pm$ 0.00 \\
t & Time on February 13 (UTC) & 02:59:13.20 $\pm$ 0.05  & 02:59:13.23 $\pm$ 0.05  & 02:59:13.21 $\pm$ 0.03 \\
\botrule
\end{tabular}}
\end{table}

\subsubsection{Asteroid shape}

Photometric measurements of 2023 CX1 were conducted as the asteroid approached Earth, between 23:30 and 02:51 UTC on the night of February 12-13. These observations produced a nearly flat light curve with variations of $\pm$0.3 magnitudes centered around an absolute magnitude of 32.7 (Fig. \ref{fig:CX1_LC}). The possible weak periodicity of 18.33 s suggested by \cite{Devogele2024} was not identified in any of our datasets. The lack of significant periodicity in the asteroid light curve, despite a high temporal resolution of $\sim$2~s, indicates that 2023 CX1 was either nearly spherical in shape or a fast rotator with a period below 2 s.

\subsubsection{Mass and size}

A complete characterization of the fireball was performed using optical and seismo-acoustic observations. In addition to the multiple photographic and video records of the fireball, the atmospheric entry of 2023 CX1 generated an infrasound signal that was detected as far as $\sim$5500 km away in Russia and as close as Flers, Normandy, within 150 km from the fall. 

Additionally, six SPLV fragments were sent for petrographic and cosmochemical analysis shortly after their recovery. These included noble gas measurements, non-destructive gamma-spectrometry (GS) and destructive Instrumental Accelerator Mass Spectrometry (IAMS). The bulk rock chemical composition of the meteorite, determined from one SPLV sample, is provided in Table \ref{tab:bulk_composition}. Photogrammetry models and laser scans, providing independent volumes measurements for four SPLV meteorites, resulted in average bulk densities of 3.353 $\pm$ 0.080 g/cm$^3$ and 3.294 $\pm$ 0.002 g/cm$^3$, respectively (see Section \ref{sec:density}). An average bulk density of 3.3 g/cm$^3$ was assumed for the rest of the analysis. 

We estimated the preatmospheric size and mass of 2023 CX1 using seven independent methods, including telescopic observations prior to impact, fireball measurements, and laboratory analyses of cosmogenic nuclides in recovered meteorites (see Section \ref{sec:preatm_size}). The initial radius estimate of 40–85 cm \cite{Devogele2024}, based on the object's absolute magnitude and assumed albedo, was further refined using optical fireball observations, energy deposition modeling from seismoacoustic data, and concentrations of cosmogenic nuclides. Each method provided constraints on the preatmospheric radius, converging toward a preferred value of 36 $\pm$ 3 cm and a total mass of 650 $\pm$ 160 kg (Table \ref{tab:mass}). 

\subsection{Asteroid behavior in Earth's atmosphere} \label{sec:fragmentation}

The fireball trajectory was determined from photographs and videos captured at 12 different sites across France, the Netherlands and the UK. The targeted high-resolution observations by the general public proved crucial for accurate trajectory determination (see Section \ref{sec:fireball_LC}). The fireball was first detected at an altitude of 101.755 km at 02:59:13.22 UTC, and tracked for 9.17 seconds down to 18.91 km.

\begin{figure}[h]
\centering
\includegraphics[width=.525\textwidth]{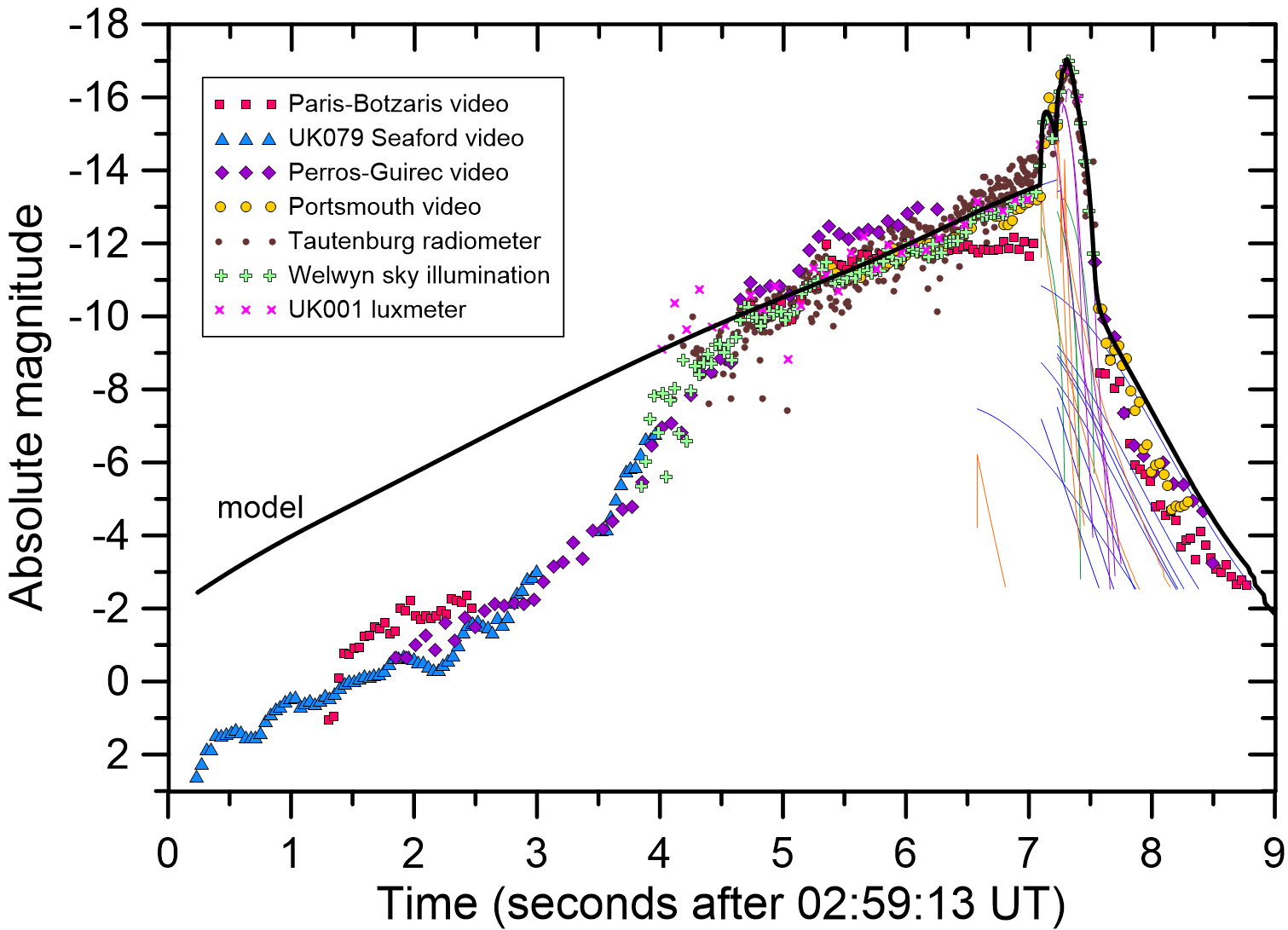}
\includegraphics[width=.46\textwidth]{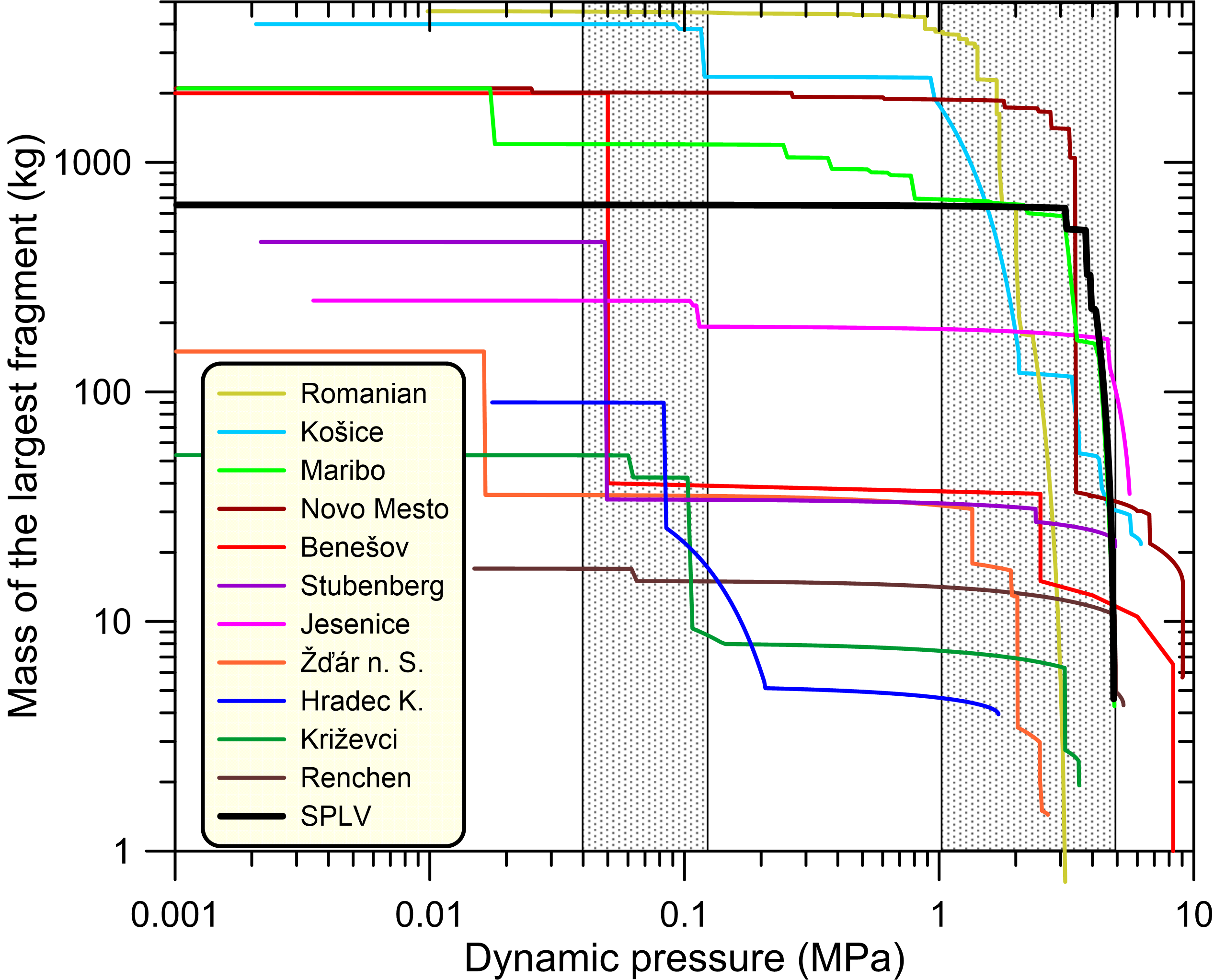}  
\caption{Left: meteor light curve expressed in visual magnitudes from a distance of 100 km. Symbols indicate data from different instruments. The modeled light curve is shown in black. Colored lines represent the individual contributions of fragments and dust released from the meteoroid, as detailed in \cite{Borovicka2020}.  Right: modeled mass of the largest surviving fragment of the SPLV meteorite as function of increasing dynamic pressure. Results are compared to the fireballs of meteorites Novo Mesto \cite{Vida2023}, Bene\v{s}ov (LL3.5-H5), Hradec Kr\'alov\'e (LL5), Stubenberg (LL6), Ko\v{s}ice (H5), Kri\v{z}evci (H6), Renchen (L5-6), Jesenice (L6) and \v{Z}d'\'{a}r nad S\'azavou (L3), Maribo (CM2) and the January 7, 2015 superbolide observed over Romania \cite{Borovicka2017,Borovicka2020}. Grey areas indicate the two distinct ranges of dynamic pressures at which dm-sized ordinary chondrites fragment \cite{Borovicka2020}.}\label{fig:fireball_LC}
\end{figure}

The fireball light curve is shown in Figure \ref{fig:fireball_LC}. Its brightness increased gradually over the first four seconds, corresponding to the body's heating-up phase before vigorous ablation starts \cite{Popova2019}. The brightness then nearly plateaued at an absolute magnitude of $-12$ during three seconds, before a bright double-peaked flare occurred, signaling two major fragmentation events at altitudes of 29.4 and 27.1 km. At the peak, the brightness increased by 3.5 magnitudes (a 25-fold increase), reaching a maximum absolute magnitude of $-17$ at 02:59:20.32 UT, corresponding to a fragmentation event under a dynamic pressure of 4~MPa. As in previous studies (e.g., \cite{Fadeenko1967,Popova2011,Popova2019,Borovicka2020}), we assume that fragmentation occurs when the dynamic pressure, which depends on both velocity and atmospheric density, exceeds the internal strength of the body.

The light curve and the deceleration of the main body and fragments were modeled using the semi-empirical fragmentation model of \cite{Borovicka2020}, assuming a meteoroid density of 3.3 g/cm$^3$ and a luminous efficiency of 4\%  (cf. Section \ref{sec:fireball_LC}). The model accurately reproduced the light curve before and after the flare, except for the pre-heating phase that was excluded from the fit. 

Both the light curve and the deceleration suggest no significant fragmentation occurred before the flare. From the bolide observations, the computed meteoroid's initial mass of 650 kg $\pm$ 160 kg (in good agreement with the other independent measurements of Table \ref{tab:mass}) decreased to 630 kg just prior to the flare. The flare itself marked the near-total destruction of 2023 CX1 (Figure \ref{fig:fireball_LC}), breaking it into small fragments and dust, which quickly evaporated. This behavior is unique among almost all other meteorite falls, which generally fragment and lose most of their mass at low dynamic pressures ($<$0.12 MPa, cf. Figure \ref{fig:fireball_LC}).

At least nine individual fragments were identified in the most detailed video recorded from the {\it La Fresnaye} village, with their estimated masses listed in Table \ref{tab:fireball_fragmentation}. Only one relatively large fragment (F), which lagged further behind the main mass, likely separated from the meteoroid before the flare, at an altitude of about 35 km and time 6.6 s (02:59:19.6 UT). Since no brightness increase was observed at that altitude, no significant dust release and only minimal mass loss occurred during the fragmentation. 

All the other individual fragments were produced during the double-peaked flare. Three large stones (A, D \& I) separated during the first fragmentation event at 29.4 km altitude, while five others were released during the second, more prominent fragmentation phase around 28.1 km. The separation of the largest fragment observed (A) at the onset of the flare caused a significant deceleration and a change in the meteoroid's direction of about $\sim$$0.8\degree$ (1.2$\degree$ southward in azimuth and 0.3$\degree$ downward in slope). The motion of fragment A was  best fitted with an elevated ablation coefficient of 0.015~$\pm$~0.003 kg MJ$^{-1}$ (three times the baseline, cf. Section \ref{sec:fireball_LC}), suggesting possible additional fragmentation between the flare and the fireball's end.

Seismoacoustic analysis of the atmospheric entry indicate that most of the body's kinetic energy was released at once during the flare. Arrival times of the resulting acoustic shock wave at seismic stations located within 150 km of the source were used to independently estimate the location of the fragmentation event (see section \ref{sec:seismo-acoustic_methods}), confirming the position derived from optical observations (see table \ref{tab:fireball_fragmentation}).

\begin{figure*}[h]%
\centering
\includegraphics[trim={0.5cm 0cm 0cm 0.5cm},clip,width=\textwidth]{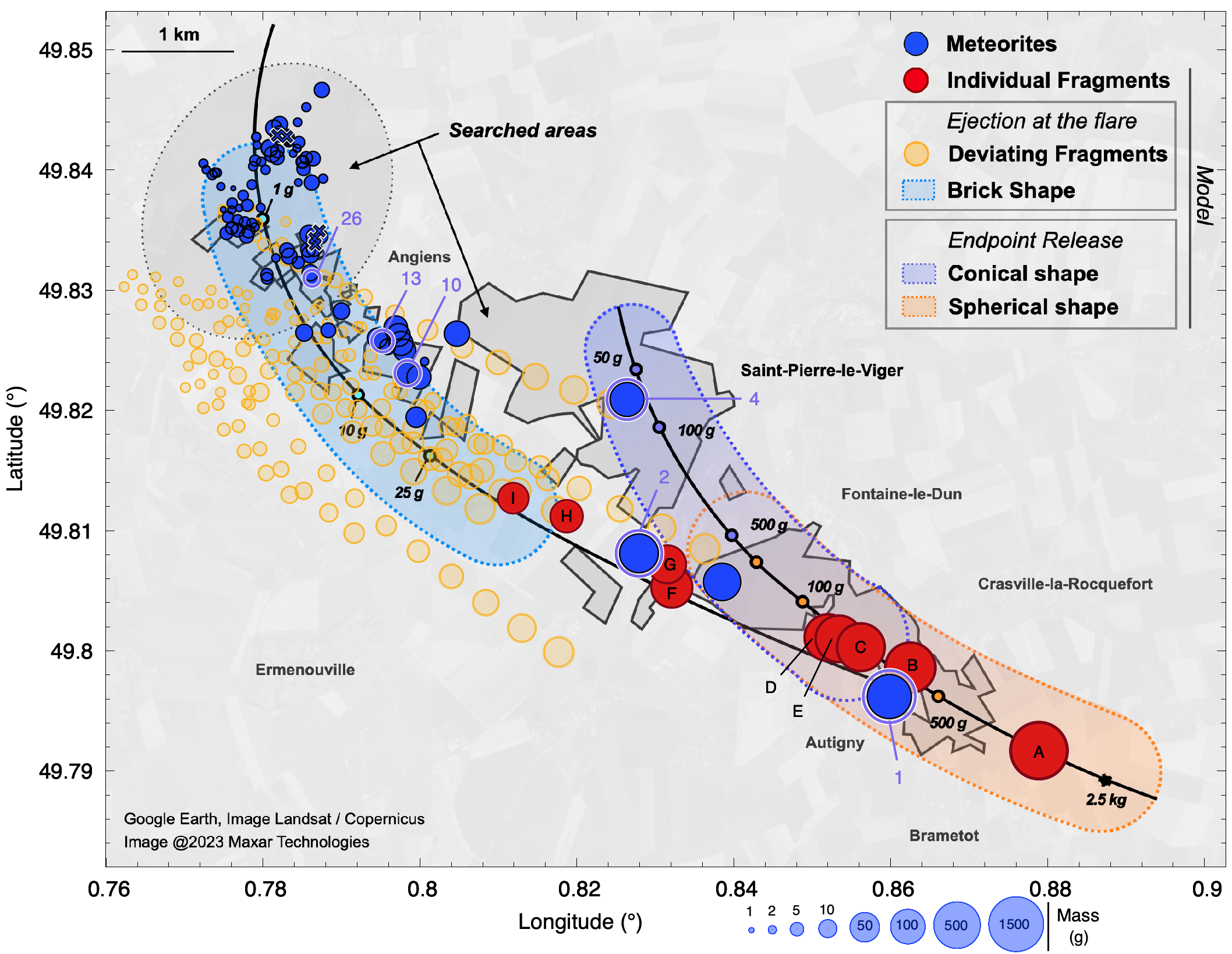}
\caption{Saint-Pierre-le-Viger meteorite strewn field map of known finds in blue circles. Meteorites with known coordinates but uncertain masses are represented by blue crosses. The positions of the stones sent for cosmogenic nuclides analysis are highlighted in purple. Red circles show the calculated fall location of the nine individual fragments identified in the video recorded from {\it La Fresnaye}. Orange circles represent the hypothetical locations of small fragments released at the main flare, with up to 5$\degree$ deviation in all directions from the original trajectory. The blue-shaded area indicates the landing sites of brick-shaped fragments weighing less than 50 g, ejected during the main fragmentation event. Orange and purple regions show the predicted landing zones for larger fragments with conical (purple) and spherical (orange) shapes, modeled from the endpoint of the visible trajectory. Circle sizes correspond to the mass of modeled or collected fragments, ranging from 1 g to 1.55 kg. The most intensively searched areas, where documented, are shaded in grey.}\label{fig:strewn_field}
\end{figure*}

\subsection{Fragment distribution on the ground}

The meteorite strewn field is presented in Figure \ref{fig:strewn_field}. 
More than 100 meteorites totaling a mass of 1.34 kg ($\sim$0.2\% of the initial mass of the meteoroid) were recovered along a distance of 8 km. All retrieved fragments showed a fusion crust, indicating that they separated from the meteoroid during ablation (i.e., while traveling at speeds exceeding $\sim$4 km/s) and did not subsequently break. The location and mass of the reported stones, ranging from $<$1 g to 490 g, are shown in Figure \ref{fig:strewn_field} and Table \ref{table:frag_coordinates}. 

Several modeling approaches were used to investigate the meteorite strewn field, with details provided in Section \ref{sec:SF_model}. Individual fragments identified in the video records after the flare were modeled to estimate the landing locations of the largest meteorites (red circles). Our results indicate that the three largest recovered stones were found near the predicted locations of fragments B, G and F, while the largest expected stone ($\sim$1.5 kg, corresponding to fragment A) has
not yet been recovered.

Stone \#4, found farther north, can be explained by a 5$\degree$ northward deviation at the main flare (orange circles in Figure \ref{fig:strewn_field}), occurring when fragment A deviated southward. Another possibility is that this meteorite separated from fragment A near the trajectory end point and experienced a strong drag afterward. However, no additional fragmentation was observed in the fireball records, hampering the confirmation of late-stage fragmentation.
Deviations of 1 to 5$\degree$ northward also reproduce the concentration of fragments recovered southeast of Angiens (near fragments \#10 and \#13). The absence of recovered meteorites in the southern part of the strewn field may be due to less thorough searches in that area.
Most of the small meteorites north of Angiens lie within 400 m of the central line (blue-shaded areas). The northernmost fragments align well with high-drag scenarios, involving ejection at the flare of elongated meteorites (see Methods \ref{sec:SF_model}).

Further insight into the fragmentation process was sought from cosmogenic nuclide studies. In well-measured past falls (e.g., \cite{Jenniskens2022}), most of the central part of the meteoroid fragments first, while the near-surface backside of the meteoroid survives longer and falls as larger fragments on the ground at the head of the strewnfield. For SPLV, however, we observe no correlation between the samples’ $^{60}$Co activity, which increases with shielding depth, and their position in the strewn field (cf. Table \ref{tab:radionuclide_activities}). Measured ($^{21}$Ne/$^{22}$Ne)$_{cos}$ ratios were used as a shielding indicator also. An upper limit for the preatmospheric radius of 50 cm (based on estimates of Table \ref{tab:mass}) was applied to better constrain the production rates of cosmogenic nuclides. Consistent ($^{21}$Ne/$^{22}$Ne)$_{cos}$ ratios were determined for all analyzed samples (cf. Table \ref{tab:CRE}), and similar shielding depths of 21-40 cm were measured for stones at the beginning (\#26), middle (\#10), and end (\#4) of the strewn field.
Variations in $^{26}$Al concentrations measured by both IAMS and GS show no significant trend along the strewn field. While the two techniques yield slightly divergent values (see Section \ref{sec:methods_IAMS}), both datasets consistently indicate no correlation between the samples' shielding depth and their position within the field. This supports the sudden disruption of 2023 CX1, that led to the dispersion of fragments from both the interior and surface of the meteoroid across the strewn field.

\subsection{Meteorite geochemistry and origin} \label{sec:meteorites_analysis}

Due to the short observation window prior to impact, no telescopic reflectance spectra could be obtained for 2023 CX1. However, the combination of its well-constrained orbit and comprehensive geochemical analyses of the recovered meteorites provides critical insight into its origin.

2023 CX1 has a Lyapunov time of 50-100 years, typical of near-Earth asteroids \cite{Whipple95}, indicating a highly chaotic orbit that limits reliable predictions beyond a few centuries. However, backward integration of clones representing the error ellipse of combined telescopic and meteor data suggests a 90\% probability that 2023 CX1 was in near-Earth space ($q<1.3$~au) 1 Ma, and an 80\% probability 10 Ma. Over the past 10 Myr, the clones spent 75\% of their time within the inner main belt ($1.3 < a < 2.5$~au), and about 1\% in the Hungaria region ($1.76<a<2.00$~au, $16<i<34\degree$), consistent with an origin as an inner main belt asteroid. 

Gas retention ages, measured by the U/Th-He (T$_4$) and K-Ar (T$_{40}$) thermochronometers, were used to trace the thermal history of the meteoroid in its parent body. These ages were remarkably consistent across the five SPLV samples analyzed (cf. Table \ref{tab:CRE}). T$_{40}$ values indicate gas retention since 4.3-4.6 Gyr for SPLV, consistent with its formation. The younger T$_4$ ages, clustering around 2.4-2.8 Gyr, suggest one or several minor resetting events (e.g., impacts) experienced by the parent body. The T$_{4,40}$ ages do not show a signature of the major parent body breakup event around 466 Myr, as recorded in $\sim$50\% of L chondrite meteorites \cite{Swindle2014}.

Cosmic ray exposure ages (CRE), used to determine the time of ejection from the parent body \cite{Herzog2014}, were also largely consistent across all SPLV samples analyzed. CRE ages determined from the $^{21}$Ne abundance, more reliable than based on the $^{3}$He and $^{38}$Ar concentrations, indicate a preferred value of 27-32 Myr for SPLV (cf. Table \ref{tab:CRE}) which is consistent with other L chondrites (e.g., \cite{Brown2023}). $T_{26}$ ages determined from the $^{26}$Al-$^{21}$Ne isotope pair method  \cite{Povinec2020} yield similar estimates around a weighted average of 30 $\pm$ 3 Myr.

SPLV's CRE age and orbital characteristics are fully consistent with the suggestion that L chondrites originate from the Massalia family, an S-type asteroid group in the inner main belt. In particular, dynamical analyses using the METEOMOD model \cite{Broz2024} suggest a $\sim$94\% probability that 2023 CX1 originated from a subpopulation of this family (identified as Massalia$_2$), formed by a cratering or reacculumation event on asteroid (20) Massalia  30–40 Ma ago \cite{Marsset2024, Broz2024}. The object was most likely delivered to Earth via the $\nu_6$ resonance (see Figure \ref{fig:meteomod_map} and Table \ref{tab:meteomod}). Following collisional simulations from \cite{Broz2024}, this family has the potential to deliver 10 to 20 meter-sized impactors to Earth every year.

\section{Discussion} \label{sec:discussion}

2023 CX1 lost approximately 98\% of its mass during a single catastrophic disruption. With fragments stopped nearly instantaneously, almost all its kinetic energy was deposited as a point source at around 28 km altitude when the dynamic pressure reached 4~MPa. Such behavior is highly atypical for common smaller meteorite-dropping fireballs, although the number of well-characterized and modeled meter-sized impactors is only $\sim$10. Most smaller chondritic meteorite-dropping fireballs  exhibit two distinct fragmentation phases: an early loss of mass at low dynamic pressures (below 0.12 MPa), sometimes followed by moderate to severe fragmentations deeper in the atmosphere (1–5 MPa) \cite{Popova2011, Borovicka2020}. 2023 CX1 did not undergo significant fragmentation at low dynamic pressures, unlike all the 22 meteorite-dropping fireballs analyzed by \cite{Borovicka2020}. 

We identified four other events in which asteroids resisted disrupting to as high pressures as 2023 CX1. 
One of the \textit{Prairie Network} fireballs studied by \cite{Ceplecha1993} survived up to 5 MPa without breaking apart, but data for this event are sparse. The Carancas impactor, an H4-5 chondrite that formed a 13-meter crater in Peru, may have reached the ground mostly intact \cite{Borovicka2008}, but no unique trajectory could be derived from its infrasound and seismic signals \cite{Brown2008}. In January 2015, a meter-sized asteroid entered over Romania and resisted pressures up to 0.9 MPa before losing $\sim$90\% of its mass through  multiple fragmentations up to 3 MPa; no meteorites were recovered \cite{Borovicka2017}. 

The most similar case to 2023 CX1 is Novo Mesto (NM), which entered over Slovenia on 28 February 2020 \cite{vida2021novo, Vida2023}. It suddenly disrupted at 35 km altitude, losing more than 80\% of its mass around 3~MPa. The meteoroid generated a 0.3 kt airburst that was detected as a minor earthquake on the ground, and dropped heavily shocked L5 chondrites. Like 2023 CX1, Novo Mesto is found to have a $\sim$96\% probability of originating from the Massalia$_2$ family, according to the METEOMOD model (cf. Table \ref{tab:meteomod}).

This fragmentation behavior is not due to the impactors being homogeneous monoliths with little or no residual damage from past collisions, as observed for the Carancas event \cite{Borovicka2008}. The dynamic pressure at breakup is at least an order of magnitude lower than that measured for recovered L chondrite meteorites \cite{Borovicka2020}. Both Novo Mesto and 2023 CX1 exhibit numerous shock veins, and computed tomography of SPLV fragments revealed macroscopic fractures likely formed by past collisional processing on the parent body.

\begin{figure*}[h]%
\centering
\includegraphics[trim={0.0cm 0.0cm 0.0cm 0.0cm},clip,width=1\textwidth]{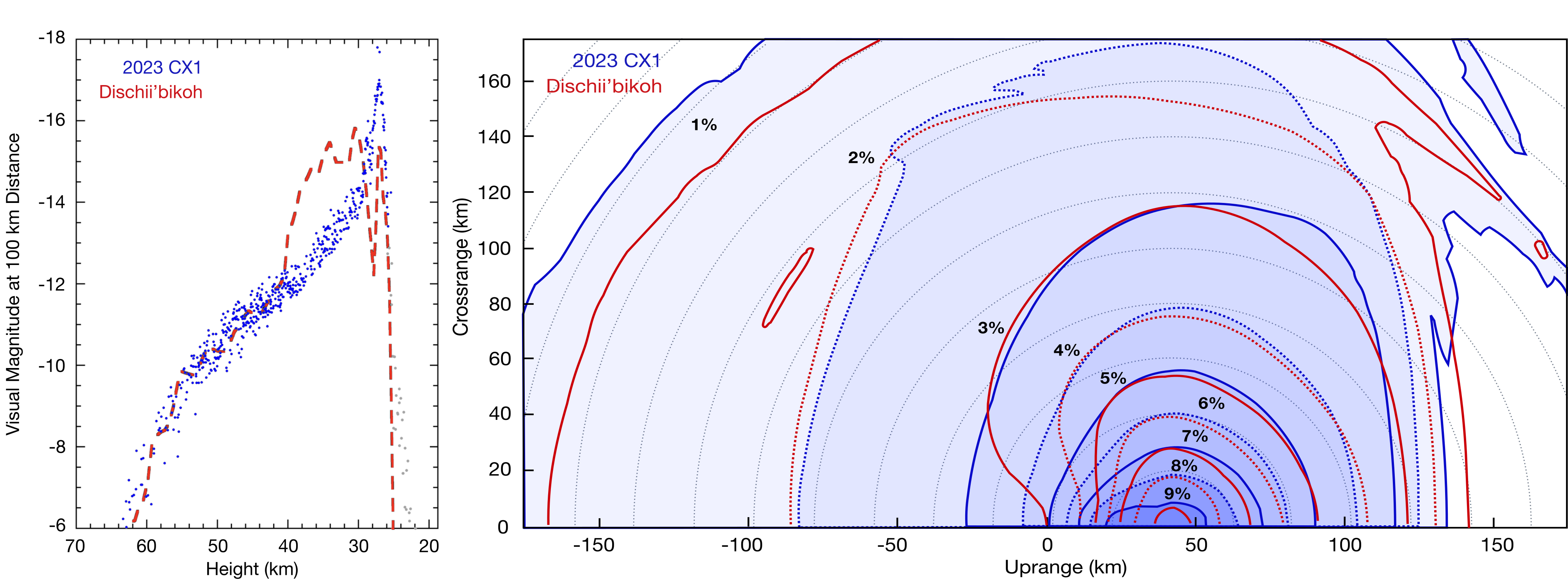}
\caption{Left: Energy deposition profile (on a logarithmic scale) of 2023 CX1 (blue) compared to the more common profile of Dishchii'bikoh (red) \cite{Jenniskens2020}. Ground-level overpressure profiles generated by a gradually fragmenting body (Dishchii'bikoh, red) and an abruptly disrupting one (2023 CX1, blue), for an impact energy of 18 Mt. Contours mark overpressure levels in percent.}
\label{fig:pressure}
\end{figure*}

\begin{figure}[ht]
    \centering
    \includegraphics[width=.9\textwidth]{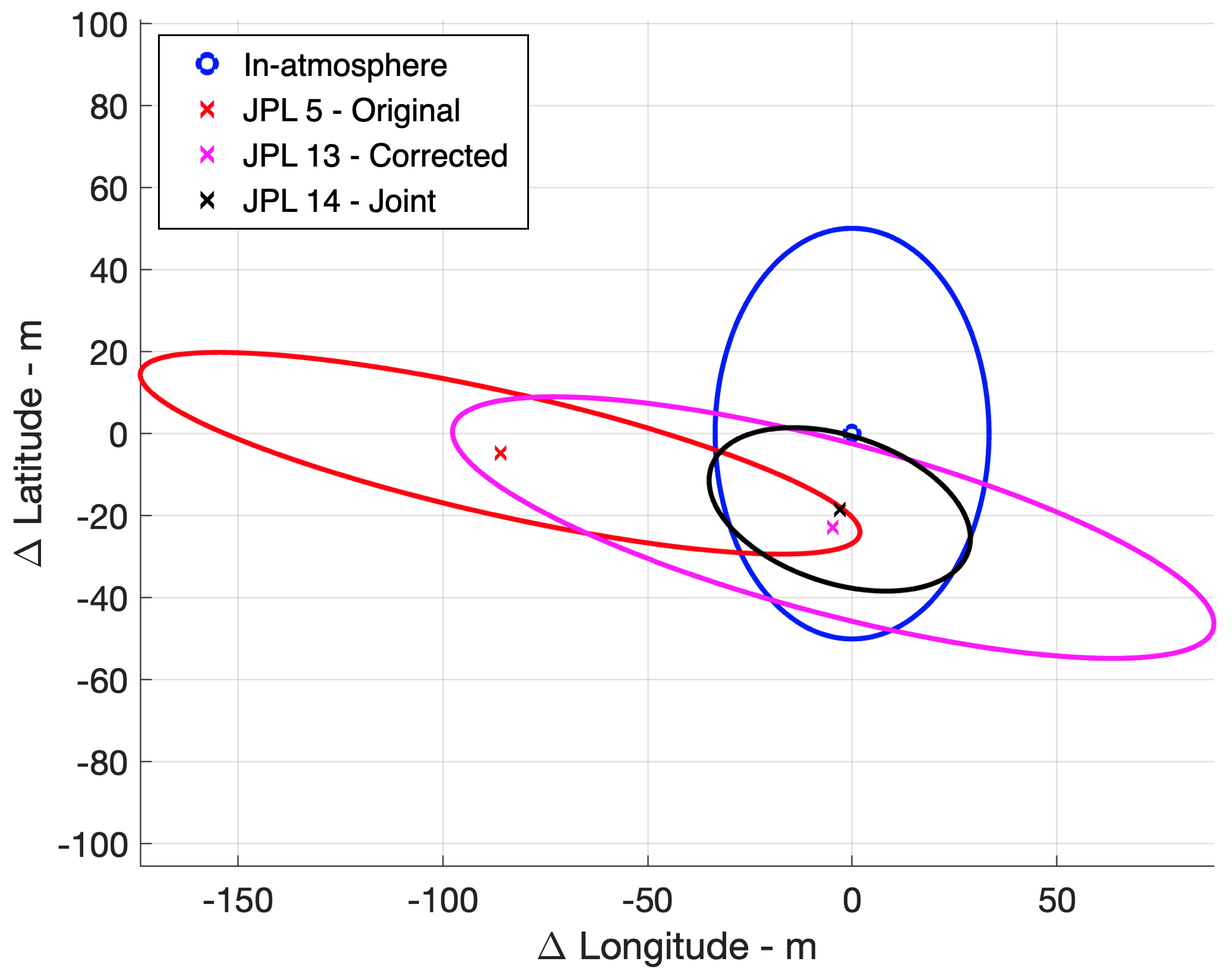}
    \caption{Comparison between the position of 2023 CX1 at the top of the atmosphere as computed from ground-based astrometry and from fireball observations (blue). The selected reference altitude of 101.755 km corresponds to the first observed position of the fireball. Ellipses represent the 3$\sigma$ confidence level. The initial solution (red) was calculated using altitudes of observing stations as originally reported to the MPC. An updated solution recalculated after correcting the observers' coordinates and assuming a 20 m positional uncertainty is shown in magenta. The joint solution (black) combines data from both telescopic and fireball observations.}
    \label{fig:orbit_comp}
\end{figure}

\begin{figure}[ht]
    \centering
    \includegraphics[width=.99\textwidth]{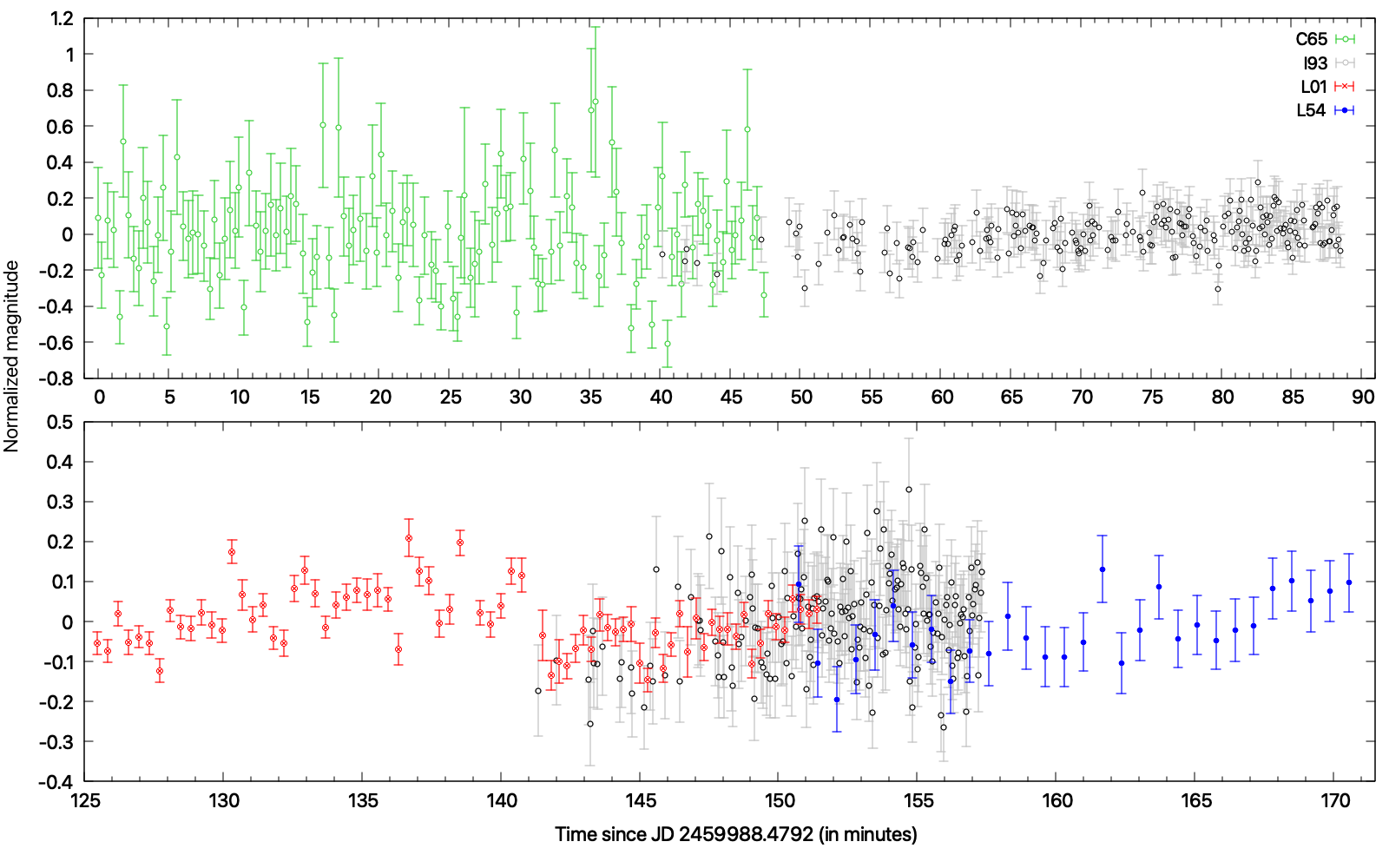}
    \caption{Asteroid brightness of 2023 CX1, starting at 23:30:02 UTC, phase angle corrected and normalized to a common distance to the asteroid, measured by the TJO (C65), St Pardon de Conques (I93), Visjan (L01) and Berthelot (L54) observatories (cf. Methods 4.2). No significant periodicity was identified in any of the datasets, including the I93 observations, which had a temporal resolution of approximately $\sim$2~s.}
    \label{fig:CX1_LC}
\end{figure}

Compared to gradually fragmenting bodies of similar size, disrupting asteroids are expected to deposit more energy closer to the ground. To investigate this effect, we compared 2023 CX1-like fragmentation to that of Dishchii’bikoh \cite{Jenniskens2020}, an LL chondrite of comparable size, entry angle, and velocity, which exhibited a more gradual fragmentation process (cf. Figure \ref{fig:pressure}). Hydrocode simulations were conducted for both fragmentation regimes, assuming an 18 Mt energy release (equivalent to the Tunguska impactor \cite{Morrison2018}), to isolate the effect of fragmentation style on shockwave propagation. While absolute overpressures scale with impact energy, the relative differences in the resulting ground footprint remain valid for lower energy events such as 2023 CX1 (see Methods \ref{sec:airburst_pressure} for details).

Significant differences in ground-level effects are observed between the two fragmentation scenarios (cf. Figure \ref{fig:pressure}). The abrupt fragmentation of SPLV generates a more spherical shockwave compared to the cylindrical shockwave in a normal entry, concentrating more energy uprange and crossrange with a deficit downrange.  As a result, the area affected by high overpressures increases substantially (by a factor of 4 at 9\%) compared with a typical entry. Differences vanish only at large distances ($>$100 km), where the source becomes effectively point-like. The closer the blast occurs to the ground, the more pronounced these differences will be, particularly at higher overpressures. Shallow entry angles will also further enhance this effect, as the cylindrical shockwave becomes more downward directed, increasing downrange impact.

These findings point to a class of L chondrite-like asteroids capable of producing disproportionately large airbursts for their size, as postulated by \cite{Vida2023}. By releasing most of their energy in a single catastrophic disruption, such objects are likely to cause greater ground-level damage than gradually fragmenting bodies of similar size (e.g., Chelyabinsk \cite{TR2021}). The fragmentation behavior observed for 2023 CX1 is consistent with the second fragmentation regime identified in other L chondrite falls \cite{Borovicka2020}, suggesting that similar responses may be expected among other damaged yet cohesive bodies from this group.

The potential for sudden and concentrated energy release among L chondrite-like asteroids, possibly linked with the Massalia family, must therefore be considered when assessing the impact threat from S- and Q-class asteroids.  In addition to the high-precision orbital tracking deployed for 2023 CX1's observation, planetary defence strategies should prioritize spectral measurements of incoming objects. Pre-impact detections pointing to a likely L chondrite-like asteroid should warrant enhanced public notification and consideration of evacuation for areas beneath the predicted fragmentation location.

\section{Methods}

\subsection{Asteroid astrometry}

Most astrometric measurements used in the trajectory determination were extracted by the original observers and reported to the Minor Planet Center (MPC). The pre-impact trajectory was based on 368  observations collected before the body entered Earth's shadow. For the first time, key observations were strategically prioritized by the European Space Agency to optimize trajectory accuracy in both spatial and temporal domains. Early night prioritization of observations from distant locations (K91 in South Africa, V39 in the United States) and the Catalina Sky Survey (703) enabled essential parallax measurements. Precise time anchoring for trajectory determination was provided by observations from Rantiga, Italy (D03), using a GPS-synchronized CMOS camera calibrated to GNSS satellites with at least 20 milliseconds accuracy, predicting impact location and timing within 400 meters and a few seconds.

Post-impact analysis involved recalibrating astrometry from multiple observatories, including Vi\v{s}njan (L01), Satino (C95), Dax (958), Conques (I93), Berthelot (L54), and  Piszk\'{e}stet\H{o} (K88), applying proper trail-fitting techniques and verifying timing and geographic coordinates. Numerous errors in the coordinates reported to the Minor Planet Center (MPC) by observers required corrections. Notably, about 55\% of the altitudes reported as above the ellipsoid (WGS84) actually referred to heights above mean sea level, with discrepancies reaching up to 95 meters.

\subsection{Asteroid photometry} \label{sec:CX1_photometry}

 Photometric reductions of the L01, L54 and I93 station data followed a similar procedure to that described in \cite{Moskovitz2019}. Standard bias and flat field corrections were applied to each data set. Photometry was then measured and calibrated using the Photometry Pipeline \cite{Mommert2017}. The Gaia DR2 reference catalog \cite{GaiaCollab2018} and Pan-STARRS DR1 catalog \cite{Flewelling2020} were used for astrometric and photometric calibrations respectively. In general, only those datasets where the signal-to-noise per exposure of 2023 CX1 was greater than $\sim$2  were selected for lightcurve analysis. All lightcurve data were collected shortly before impact when the viewing geometry to 2023 CX1 was changing significantly. The photometry was thus range (heliocentric and geocentric) and phase angle corrected. These corrections were applied assuming a standard H-G photometric model \cite{Bowell1989}, an absolute magnitude H=32.76, and geometry parameters retrieved from the JPL Horizons system. Photometric reductions of the K88 station data were performed using the tasks of FITSH software \cite{Pal2012}. C65 observations were conducted over a 45-minute interval beginning at 23:30 UT on 12 February 2023, employing continuous 5-second exposures through a Johnson-Cousins R filter. The images were reduced using the ICAT analysis tool \cite{Colome2010}. Finally, high-resolution temporal measurements of the asteroid's magnitude variations were obtained from observations at the Schiaparelli Observatory (204), with an effective exposure time of 0.25 seconds during 60-second observation windows. Photometric reductions were performed using the Astrometrica software \cite{Herbert2012}. Searches for periodicity in the asteroid’s brightness variations were conducted across datasets L01, L54, C65, I93, K88, and 204, but no reliable signature of the asteroid’s rotation was identified in any of them.

\subsection{Preatmospheric radius and mass} \label{sec:preatm_size}

The preatmospheric radius and mass of 2023 CX1 were independently determined from asteroid photometry, optical and seismoacoustic analysis of the fireball, and cosmogenic nuclide measurements of SPLV samples. These estimates, summarized in Table \ref{tab:mass},  assume a spherical shape and an average bulk density of 3300 kg/m$^3$ measured for SPLV. Geometric albedos were computed from the asteroid's absolute magnitude and estimated size using the formalism of \cite{Bowell1989}. 

Initial radius estimates based on the asteroid’s absolute magnitude of 32.7$\pm$0.3 ranged from 40 to 85 cm \cite{Devogele2024}. Unfortunately, no reflectance spectra or direct albedo measurements were obtained prior to impact, preventing a firm constraint on the body’s size from photometry alone. However, L-chondrites have long been linked to S-type asteroids \cite{Gaffey1993,Nakamura2011}; in particular, 2023 CX1 is suspected to originate from the Massalia family (see Section  \ref{sec:meteorites_analysis}), whose members exhibit L-chondrite-like compositions \cite{Marsset2024}. From the Virtual Observatory Solar System Open Database Network \cite{Berthier2023}, we derived a geometric albedo of 0.196$\pm$0.036 for (20) Massalia, consistent with typical S-type asteroid values (e.g., 0.258$\pm$0.087 \cite{DeMeo2013} and 0.208$\pm$0.079 \cite{Usui2013}). This refined albedo constrains the asteroid's preatmospheric radius, based on brightness alone, to 35-55 cm \cite{Bowell1989}. 

Cosmogenic noble gas measurements of five SPLV samples show a pronounced excess of $^{80}$Kr and, less well-resolved, of $^{82}$Kr that are likely due to neutron capture on $^{79}$Br and $^{81}$Br. Such neutron-induced isotopes are produced in greater depths when the secondary neutrons have slowed down to enable efficient capture, indicating that 2023 CX1 was at least 30 cm in radius \cite{Eberhardt1963}. 

Gamma-spectrometry of SPLV samples further constrained the meteoroid’s size by measuring the activities of short-lived radionuclides (SLRs) such as $^{22}$Na, $^{54}$Mn, $^{57}$Co and $^{60}$Co (cf. Table \ref{tab:radionuclide_activities}. The low abundance of $^{60}$Co suggests the meteoroid was too small to sustain a full nucleonic cascade within its interior \cite{Wieler1996}. The preatmospheric radius was estimated from the SLRs concentrations, combined the meteorite's bulk chemical composition of (Table \ref{tab:bulk_composition}) and Monte Carlo simulations described in  \cite{Wieler1996,Leya2009,Leya2021}. The measured concentrations of $^{60}$Co and $^{26}$Al yielded radius estimates of 25 $\pm$ 4 cm and 28 $\pm$ 12 cm, respectively. 
Further constraints were provided by Instrumental Accelerator Mass Spectrometry (IAMS) analysis. Measured $^{26}$Al activity in SPLV sample \#1 was interpreted using simulations of \cite{Leya2021}, yielding a minimum radius of 20 cm (cf. Table \ref{tab:IAMS}).

The atmospheric entry of 2023 CX1 generated an infrasound signal that was recorded by multiple arrays located in Europe, North Africa and Russia. From the observed wave periods, the overall energy released during the event was 0.029 $^{+0.026}_{-0.014}$ kt of TNT equivalent (1.2$^{+1.1}_{-0.6}$ $\times$ 10$^{11}$ J, see section \ref{sec:infrasound}). This corresponds to an initial mass of 1230$^{+2300}_{-630}$ kg and a radius of 36-55 cm, consistent with telescopic estimates.

\subsection{Meteor optical analysis} \label{sec:fireball_LC}

The fireball trajectory and velocity were determined using the method of \cite{Borovicka1990,Borovicka2022}. Cameras from the Global Meteor Network (GMN) in the UK where the skies were clear captured the initial phase of the fireball. Absolute timing was obtained from a radiometer of an European Fireball Network camera in Tautenburg, Germany. Mobile phone recordings taken close to the fireball's end, several of which were hand-held, enabled the determination of the fireball's middle and end segments. Photometry was done on one GMN camera, one mobile phone, one security camera and another low-sensitivity record from a mobile phone showing no stars but the Moon. In addition, the bright part was covered by one all-sky video under overcast sky and one calibrated luxmeter under overcast sky (both in the UK) and the radiometer in Tautenburg, with low signal-to-noise but high temporal resolution. 

The trajectory computation was performed following \cite{Borovicka1990}, and independently confirmed using the method of \cite{Vida2019}. No single straight-line trajectory solution could reproduce the meteoroid’s motion both before and after the major fragmentation event at 27–29 km altitude, due to a measurable change in direction following the flare. Accurate determination of the trajectory therefore required fitting two separate solutions to the observations, one before and one after the fragmentation. The resulting trajectory parameters are summarized in Table \ref{tab:trajectory}.

The light curve and deceleration of the main body and individual fragments were fitted by the semi-empirical fragmentation model of \cite{Borovicka2020}. The same 4\% luminous efficiency was used for both big and small fragments. The usually used values are 5 \% for big and 2.5 \% for small fragments \cite{Borovicka2020}. The drag and shape coefficients were set to $\Gamma A$ = 0.7 and the density was set to 3300 kg m$^{-3}$. 

For SPLV, an ablation coefficient $\sigma$ of 0.003 kg MJ$^{-1}$ was found to best reproduce the light curve prior to the flare, consistent with pure thermal ablation in the absence of fragmentation \cite{Ceplecha2005}. Due to limited observational coverage after the flare, the ablation coefficients of individual fragments could not be rigorously constrained. A standard value of $0.005$~kg~MJ$^{-1}$ \cite{Borovicka2020} was assumed for all fragments except one and was found to be consistent with the available data. The only exception was fragment A, whose dynamics required an elevated $\sigma$ of 0.015~$\pm$~0.003~kg~MJ$^{-1}$ to adequately match the observed deceleration, indicating additional fragmentation along the trajectory. The dynamics of fragment A could alternatively be fitted using a standard $\sigma$ of 0.005~kg~MJ$^{-1}$ combined with one or two fragmentation events, although the available observations do not allow for robust constraints on the timing and mass change associated with such events. 

\subsection{Strewn field modelling} \label{sec:SF_model}

Independent modeling approaches were combined to analyze the meteorite strewn field. The landing positions of the nine individual fragments observed in video recordings were predicted using dark flight (DF) computations based on the ALADIN (\quotes{Aire Limitée Adaptation dynamique Développement InterNational}) atmospheric wind model at 03:00 UTC. Meteorological wind models were provided without associated uncertainties. Additional DF computations using alternative wind profiles (e.g., radiosonde data from the Herstmonceux area) occasionally resulted in a slight overall shift in the strewn field position but did not significantly affect the modeled dispersion of the meteorites. For the main analysis, we used the wind model that provided the best agreement with the locations of the recovered meteorites.

To explain the concentration of meteorites recovered southeast of Angiens, we simulated the dark flight of hypothetical spherical fragments ranging from 1 to 66 grams, released during the flare with varying trajectory deviations. For each significant fragmentation or dust release event, five representative fragments were modeled—one following the original trajectory and four deviating by up to five degrees in the up, down, left, and right directions. Each underwent ablation and dark flight simulation to determine its impact point (see orange circles in Figure \ref{fig:strewn_field}). The limit five-degree deviation was chosen to match the observed displacement of the most off-axis recovered meteorite (\#4), although it should be assumed that deviations follow a normal distribution with an outer edge around five degrees. Although large, such deviations are plausible given the 0.8$\degree$ offset measured for the most massive fragment (A) after the flare. This behavior may reflect an explosive fragmentation process, possibly involving chemical reactions such as magnesium oxidation. While simplified, this approach offers a useful approximation of how lateral dispersion may have shaped the final strewn field. 

Additional DF models were conducted for fragments of different masses and shapes, released at various altitudes, following \cite{Vida2023}. First, a DF model was applied to small ($<$50 g) fragments ejected at the main flare, accounting for shape variations (conical, brick, and spherical) to evaluate drag effects. For SPLV, higher drag (e.g., conical shapes) shifted the predicted strewn field north along the central line, while lower drag (spherical shapes) shifted it southeast. A scenario using an intermediate shape (brick) is shown in Figure \ref{fig:strewn_field} as a blue-shaded area. 
Another limiting case, assuming larger conical and spherical fragments released at the end of the fireball’s visible path, is illustrated in Figure \ref{fig:strewn_field} as orange and purple shaded areas. 

\subsection{Meteor infrasound analysis} \label{sec:infrasound}

The atmospheric entry of 2023 CX1 generated an infrasound signal that was recorded by several stations of the Comprehensive Nuclear-Test-Ban Treaty Organization (CTBTO) and by several arrays located in the Netherlands (KNMI), the United Kingdom and France (CEA), including at a station located within 150 km from the fall in Flers, Normandy. For all International Monitoring System stations (with station labels starting with \quotes{I} in Table \ref{tab:infrasound}), the procedure for bolide infrasound analysis described in \cite{Edwards2006, Ens2012, Gi2017} was followed. The characteristics of the infrasound signals recorded by each station are detailed in Table \ref{tab:infrasound}. To estimate the source energy, the empirical relation between multi-station period averages and yield given in \cite{Ens2012} was used (using the period at maximum amplitude zero crossing method by \cite{ReVelle1997} and the peak in the bolide infrasound signal power spectral density \cite{Ens2012}).

\subsection{Meteor seismo-acoustic analysis} \label{sec:seismo-acoustic_methods}

The seismo-acoustic wave was recorded by several seismic sensors in northern France and southern UK which are part of the \quotes{Réseau Sismologique et géodésique Français} (RESIF) French permanent broadband seismic network \cite{RESIF} and the Great Britain Seismograph Network, respectively. The waves were also recorded by the \textit{Raspberry Pi Shake \& Boom} citizen science initiative, including four stations located within 150 km from the event (cf. Table \ref{tab:seismo_acoustic} and Figure \ref{fig:3D_acoustic}). Arrival times were evaluated using atmospheric models from the European Centre for Medium-Range Weather Forecasts (ECMWF IFS cycle 38r2), and 3D-ray tracing simulations calculated with WASP-3D \cite{Virieux2004}. One eigen ray is identified for each arrival and celerity models are extracted by dividing the 3D distances in Cartesian coordinates by travel times. 

To account for the effects of unresolved gravity waves in the used atmospheric models \cite{Listowski2024} on the propagation, errors of the predicted arrival times were estimated by adding one thousand random perturbations uniformly distributed throughout $\pm$ 10 m/s to the 3D celerity models. For each realization, a 3D location was performed using a non-linear least squares solver \cite{More2006} which optimizes the latitude, longitude, origin time and altitude of the fragmentation from the minimization of the RMS (Root Mean Square) time residuals. The best agreement between the theoretical and observed arrival times was reached for a source located at latitude 49.81 $\pm$ 0.05$\degree$ N, longitude 0.64 $\pm$ 0.07$\degree$ E and a height of 27.9 $\pm$ 1.0 km, in agreement with optical data (see table \ref{tab:fireball_fragmentation}).

\subsection{Contribution of citizen science networks} \label{sec:citizen_network}

Citizen contributions were crucial at every stage of the 2023 CX1 analysis. Shortly after its discovery, amateur astronomers worldwide performed numerous follow-up observations, enabling refining the asteroid’s orbit. Timely social media communication led to the first large-scale targeted observations of a fireball in the atmosphere. Still images and videos recorded using hand-held phones (e.g., from the \textit{La Fresnaye} city) were essential for computing the bolide’s trajectory. Observers also recorded calibration images with their devices to aid trajectory computation. Several cameras of the Global Meteor Network hosted by amateur astronomers caught the fireball. In addition to professional infrasound and seismic measurements, the citizen project Raspberry Pi \& Boom network acquired and analyzed seismic data, contributing to the analysis of 2023 CX1’s fragmentation. Finally, the rapid mobilization of the public in Normandy was crucial to the recovery of meteorites. Half of the meteorites collected through the FRIPON/Vigie-Ciel citizen science project were found by volunteers. All meteorites retrieved by the Vigie-Ciel team were collected responsibly and in accordance with local legislation.

\subsection{Meteorite density}  \label{sec:density}

Overlapping images were taken from various angles to document the full exterior of four fragments of different sizes. Structure-from-Motion (SfM) photogrammetry was used to process these images to reconstruct 3D shape models in the software Agisoft Metashape Professional (v.2.0.2). In various steps that involve (1.) camera matching, the generation of (2.) a sparse cloud, (3.) a dense cloud, and scale bars used during the photo-documentation procedure, the software is capable of rendering true-size 3D shape models. Additional 3D models were also independently obtained through laser scanning. Estimates of the volumes of the generated 3D shape models were then determined with the Meshlab (v2022.02) software and used with the fragment's mass to find their bulk rock densities (see Table \ref{tab:density}). 

\subsection{Meteorite noble gas mass spectrometry} \label{sec:methods_noble_gas}

Cosmogenic nuclides are produced by the irradiation of surface materials by primary cosmic-ray protons of both galactic and solar origin. They serve as valuable tracers for investigating the origin and characteristics of meteorites, including the meteoroid's preatmospheric size, cosmic-ray exposure age, and the shielding depth of individual fragments. To reconstruct the exposure history of 2023~CX1, we performed noble gas analyses on five SPLV meteorite fragments recovered across the strewn field.

All stable He-Xe isotopes were measured at ETH Zurich using the inhouse-built noble gas mass spectrometer Albatros with established procedures described by \cite{Riebe2017}. One aliquot each was measured from SPLV fragments \#1, 4, 10, 13, and 26. The sample fragments ($\sim$16-23 mg, Table \ref{tab:CRE}) were wrapped in Al foil and pre-heated at $\sim$110$\degree$C in ultra-high vacuum for several days prior to analysis to release adsorbed atmospheric gases. The noble gases were extracted by fusion in a Mo-crucible at $\sim$1700$\degree$C for $\sim$25 min. A re-extraction at $\sim$1750$\degree$C performed for SPLV \#13 verified full gas extraction (gas amounts were $<$ 1\% of the main step for all elements). Blank corrections were $<$ 0.3\% of the signals for all He and Ne isotopes, $<$ 4.4\% for $^{38}$Ar and $^{40}$Ar, $<$ 16\% for $^{36}$Ar and all Kr isotopes, and $<$ 3\% for all Xe isotopes.

Noble gas measurements revealed that the He, Ne, and Ar isotopic compositions of SPLV samples are mostly cosmogenic (cos, see Table \ref{tab:noble_gas_concentrations}). No solar wind component was detected, indicating that SPLV is not a regolith breccia. Radiogenic (rad) $^4$He and $^{40}$Ar were also identified, with a trace of trapped (tr) Ar. The elemental ratios of the heavy noble gases as well as the $^{130,131}$Xe/$^{132}$Xe ratios do not reveal significant air contribution, reflecting the fast recovery of SPLV. The $^{129}$Xe/$^{132}$Xe ratios elevated over the chondritic Xe indicate minor amounts of short-lived radionuclide $^{129}$I-derived $^{129}$Xe in all samples, suggesting the incorporation of $^{129}$I into SPLV early in the solar system (see Table \ref{tab:noble_gas_concentrations}).

The contribution from cosmogenic nuclides was derived as follows. All samples lack evidence for a Ne component. Thus, He and Ne in all samples are purely cosmogenic, apart from $^{4}$He (Table \ref{tab:CRE}). The $^{36}$Ar/$^{38}$Ar ratios measured for all samples are slightly higher ($\sim$0.98-1.53) than typical cosmogenic values (0.63-0.67; \cite{Wieler2002}) indicating the presence of minor Ar$_{tr}$, in addition to dominant Ar$_{cos}$ and $^{40}$Ar$_{rad}$. A two-component deconvolution was performed between typical ($^{36}$Ar/$^{38}$Ar)$_{cos}$ and ($^{36}$Ar/$^{38}$Ar)$_{tr}$ of 5.32-5.34 (covering Q and air compositions) to determine $^{38}$Ar$_{cos}$ (Table \ref{tab:CRE}) in these samples. ($^{80}$Kr/$^{82}$Kr)$_\text{excess}$ ratios were determined by assuming $^{86}$Kr to be entirely trapped, ($^{80-83}$Kr/$^{86}$Kr)$_\text{tr}$ covering Q and air, and applying a typical ($^{80,82}$Kr/$^{83}$Kr)$_\text{cos}$ ratio for ordinary chondrites (0.487-0.585/0.705-0.771; \cite{Leya2015}). We obtained ($^{80}$Kr/$^{82}$Kr)$_\text{excess}$ ratios (Table \ref{tab:CRE}), that are, within error, in a typical range expected for neutron capture derived Kr (\cite{Martin1966}), confirming that the $^{80,82}$Kr excesses are likely due to neutron capture on $^{79,81}$Br.

The CRE age was determined as follows. We used the ordinary chondrite matrix model by \cite{Leya2009} and the bulk chemical composition of SPLV \#1 analysed by ICP-AES/ICP-MS in this study (Table \ref{tab:bulk_composition}) to determine the production rates for cosmogenic $^{3}$He, $^{21}$Ne, and $^{38}$Ar and the respective cosmic ray exposure (CRE) ages $T_x$ for each sample. Each cosmogenic nuclide provides CRE ages that are similar for all the samples analysed, except for a slightly higher $T_{38}$ for fragment \#10 and a smaller $T_{38}$ for fragment \#1. This small deviation is likely caused by target element heterogeneities (the so-called \quotes{nugget effect}) that predominantly influence the $^{38}$Ar$_{cos}$ concentrations. We observe a tendency of lower CRE ages with lower mass of the respective cosmogenic isotope, i.e., $T_{3} < T_{21} \leq T_{38}$. This may indicate a loss of noble gases due to heating e.g., during atmospheric entry or passage close to the Sun, which preferentially affects the lighter noble gases. However, the SPLV data plot well on the correlation line in the “Bern plot” (cf. Fig. 3 in \cite{Nishiizumi1980}) suggesting no significant $^3$He loss. Potentially, the deviations between various $T_x$ could also result from slightly mismatching calculations of the production rates in the applied model. Lowering $P_3$ according to \cite{Dalcher2013}, who reported differences between modelled and measured chondrite data, would bring the $T_{3}$ for SPLV into agreement with the $T_{21}$  and $T_{38}$. We consider the $T_{21}$ ages most reliable, as $^{21}$Ne$_{cos}$ is less affected by heating and target element heterogeneities compared to cosmogenic $^3$He and $^{38}$Ar, respectively. 

The $^{4}$He$_{rad}$ concentrations were calculated by utilizing that $^{3}$He is entirely cosmogenic, ($^{4}$He/$^{3}$He)$_{cos}$ = 5.2-6.1 (\cite{Wieler2002}), and $^{4}$He$_{tr}$ is negligible. The $^{40}$Ar$_{rad}$ concentrations were derived by determining $^{40}$Ar$_{tr}$ based on $^{36}$Ar$_{tr}$ (see deconvolution above) and ($^{40}$Ar/$^{36}$Ar)$_{tr}$ = 0-295.5 (conservatively covering Q, with no trapped $^{40}$Ar, and air composition). The slightly lower T$_4$ age of SPLV \#4 might be explained by variations in the phosphate (apatite and merrillite) abundance ($\sim$0.5-0.7\% in L chondrites; \cite{Lewis2016}) across the different SPLV fragments that mainly controls the U concentration in the material. 

\subsection{Meteorite gamma-spectrometry (GS)} \label{sec:methods_gamma_spec}

Fragments \#2, \#4, and \#10 were also analyzed for cosmogenic radionuclides using nondestructive gamma-spectrometry at the low-background laboratory of Comenius University in Bratislava. This gamma-spectrometry (GS) analysis provided an independent constraint on the impactor's preatmospheric radius and the shielding depths of the meteorites, as well as a complementary estimate of the cosmic-ray exposure age of 2023 CX1. 

Two high-purity germanium (HPGe) detectors (Princeton Gamma Tech (PGT), USA) and (Canberra/Mirion, USA) of 70\% and of 50\% relative efficiency (for 1332.5 keV gamma-rays of $^{60}$Co), respectively, placed in low-background shields situated in the basement of the three-story building were used for analyses. A detailed description of the gamma-ray spectrometers, measuring procedures, Monte Carlo detector's efficiency, self-absorption and summation effects corrections calculations are given in \cite{Povinec2020b,Kovacik2013}. The uncertainties of results were mainly due to counting statistics, which were typically below 10\%. The measuring time was from 30 to 50 days, depending on the mass of the analyzed samples. All measurements were corrected back to the date of the meteorite fall on 13 February 2023. Regular analysis of International Atomic Energy Agency (IAEA) and National Institute of Standards and Technology (NIST) reference materials and participation in intercomparison exercises guarantee to maintain a good quality of results. 

The $T_{26}$ cosmic ray exposure (CRE) ages presented in Table \ref{tab:CRE} were estimated using the $^{26}$Al-$^{21}$Ne isotope pair method, as described in \cite{Povinec2020,Eugster2002}, based on the $^{21}$Ne measurements in the same table. We find the $T_{26}$ ages to be consistent with estimates obtained from single noble gas isotope methods, particularly with those derived from the $^{21}$Ne method (see Table \ref{tab:CRE}).

\subsection{Instrumental Accelerator Mass Spectrometry (IAMS)} \label{sec:methods_IAMS}

Fragments \#1, \#4, \#10, \#13, and \#26 were further analyzed using Accelerator Mass Spectrometry (AMS) at the Vienna Environmental Research Accelerator (VERA). Measurements of long-lived cosmogenic radionuclides in the SPLV samples provide key constraints on the meteoroid’s preatmospheric size and the shielding depths of the recovered fragments, offering valuable insights into the distribution of material during atmospheric fragmentation. 

A distinctive innovation of VERA is an ion-laser interaction mass spectrometry (ILIAMS) system that enables isobar suppression by up to 14 orders of magnitude \cite{Martschini2022}. Consequently, the combination of ILIAMS and AMS, so-called instrumental AMS (IAMS), allowed the direct detection of $^{26}$Al/$^{27}$Al and $^{41}$Ca/$^{40}$Ca in crushed SPLV samples containing $\sim$ 1\% intrinsic Al and Ca. The presence of isobars from the natively abundant elements (15\% Mg, $\sim$0.1\%~K) does not present any analytical issues, thereby rendering radiochemical separation unnecessary.

IAMS was conducted on small aliquots of homogenized powder of $\sim$1~g of the \#1 specimen, which was also used for the determination of the bulk chemistry (Table \ref{tab:bulk_composition}), 245 mg of \#4, 233 mg of \#10, 226~mg of \#13, and 214~mg of \#26, respectively. For the extraction of AlO$^{-}$ and the measurement of $^{26}$Al/$^{27}$Al \cite{Lachner2021}, $\sim$5~mg of the fine-grained powder was pressed in Cu cathodes without the use of a metal binder. In order to increase CaF$_3^-$ extraction from the Cs ion sputter source, 0.9-1.4~mg of each SPLV powder sample was mixed with a factor of 9 by weight of PbF$_2$. This mixture was then pressed in Cu cathodes. In-house standards \quotes{Dhurmsala} ($^{26}$Al/$^{27}$Al=(1.287 $\pm$ 0.034)$\times 10^{-10}$) and SMD-Ca-11 ($^{41}$Ca/$^{40}$Ca = (0.9944 $\pm$ 0.0092)$\times 10^{-11}$), which are traceable to primary standards \cite{Rugel2016}, have been used for the purpose of normalization.

For conversion of nuclide ratios into massic activities in disintegrations per minute (dpm) per kg (Table \ref{tab:IAMS}), the Al and Ca concentrations of Table \ref{tab:bulk_composition} (1.19\% Al; 1.19\% Ca) have been used. These massic activities measured by IAMS are compared with Monte-Carlo calculation-based radius- and depth-dependent production rates for $^{26}$Al \cite{Leya2021} and $^{41}$Ca \cite{Leya2009}. The theoretical production rates are generally based on L chondrite parameters and have been adjusted to the bulk composition of SPLV in Table \ref{tab:bulk_composition}, an oxygen content of 38.47\%, and the sulfur value of 2.22\% value from \cite{Cripe1975}. Oxygen, carbon (0.08\% \cite{Grady1989}) and nitrogen (1~$\mu$g/g, \cite{Hashizume1995}) do not produce any $^{26}$Al or $^{41}$Ca, but in this way, all input elements including all relevant minor elements determined add up to 100\%. 

IAMS determined nuclide ratios can be explained by sufficiently long CRE ages (see Table \ref{tab:CRE}), resulting in saturation of $^{26}$Al and $^{41}$Ca in the samples. The reported $^{26}$Al data have total 1-sigma uncertainties of 3.1-4.0\%, which include counting statistics, sample scatter, variability in standard measurements (0.85\%), and the nominal reference value uncertainty, all combined quadratically. The massic activity uncertainty in dpm/kg also incorporates a 5\% uncertainty for stable $^{27}$Al, making it the largest contributor. Similarly, the $^{41}$Ca/$^{40}$Ca data have 1-sigma uncertainties of 14-22\%, accounting for counting statistics, sample scatter, variability in standard measurements, and the nominal reference value uncertainty. For massic activity (in dpm/kg), the additional 5\% uncertainty for stable Ca (1.19\%) contributes minimally. 

Measurements of $^{26}$Al concentrations using IAMS (Table \ref{tab:IAMS}) show a discrepancy compared to values obtained by GS (Table \ref{tab:radionuclide_activities}) across all measured SPLV samples. The lower $^{26}$Al concentrations reported by IAMS suggest shallower shielding depths (ranging from 2.5 to 30 cm; see Table \ref{tab:IAMS}) than those previously estimated. While the cause of this discrepancy remains under investigation, both IAMS and GS data confirm the absence of any correlation between shielding depth and sample position within the strewn field.

To our knowledge, the IAMS-measured $^{41}$Ca/$^{40}$Ca ratios for the five SPLV samples represent the fifth $^{41}$Ca depth profile ever obtained in a meteorite, and the first in an ordinary chondrite. The variations in these ratios are relatively narrow, ranging from 2.6 to 4.0$\times10^{-12}$. When comparing these ratios with experimental depth-dependent IAMS $^{41}$Ca/$^{40}$Ca data from the L5 chondrite Knyahinya (unpublished; shielding positions and radius of 45 cm from \cite{Graf1990}), shallow shielding depths ($<$20 cm) can be constrained. However, no further conclusions on the radius or shielding depths of the SPLV samples can currently be drawn from the comparison to production rates from Monte Carlo simulations (e.g., as noted by \cite{Bischoff2024}). 

\subsection{Asteroid dynamical analysis} \label{sec:bck_int}

The orbital stability of 2023 CX1 was analyzed through backward integration of its nominal orbit and 99 clones, generated from the orbital solution and covariance matrix of solution \#12 in Table \ref{tab:orbit}. Using a RADAU 15 integrator with a tolerance of $10^{-12}$ and a one-day external time step, we considered the gravitational influences of the Sun, the eight planets, and the Moon, based on the DE405 Planetary ephemeris \cite{Standish1998}.

\begin{figure}[!ht]
\centering
\begin{tabular}{c}
\includegraphics[trim={0.8cm 5.5cm 0cm 1cm},clip,width=.9\textwidth]{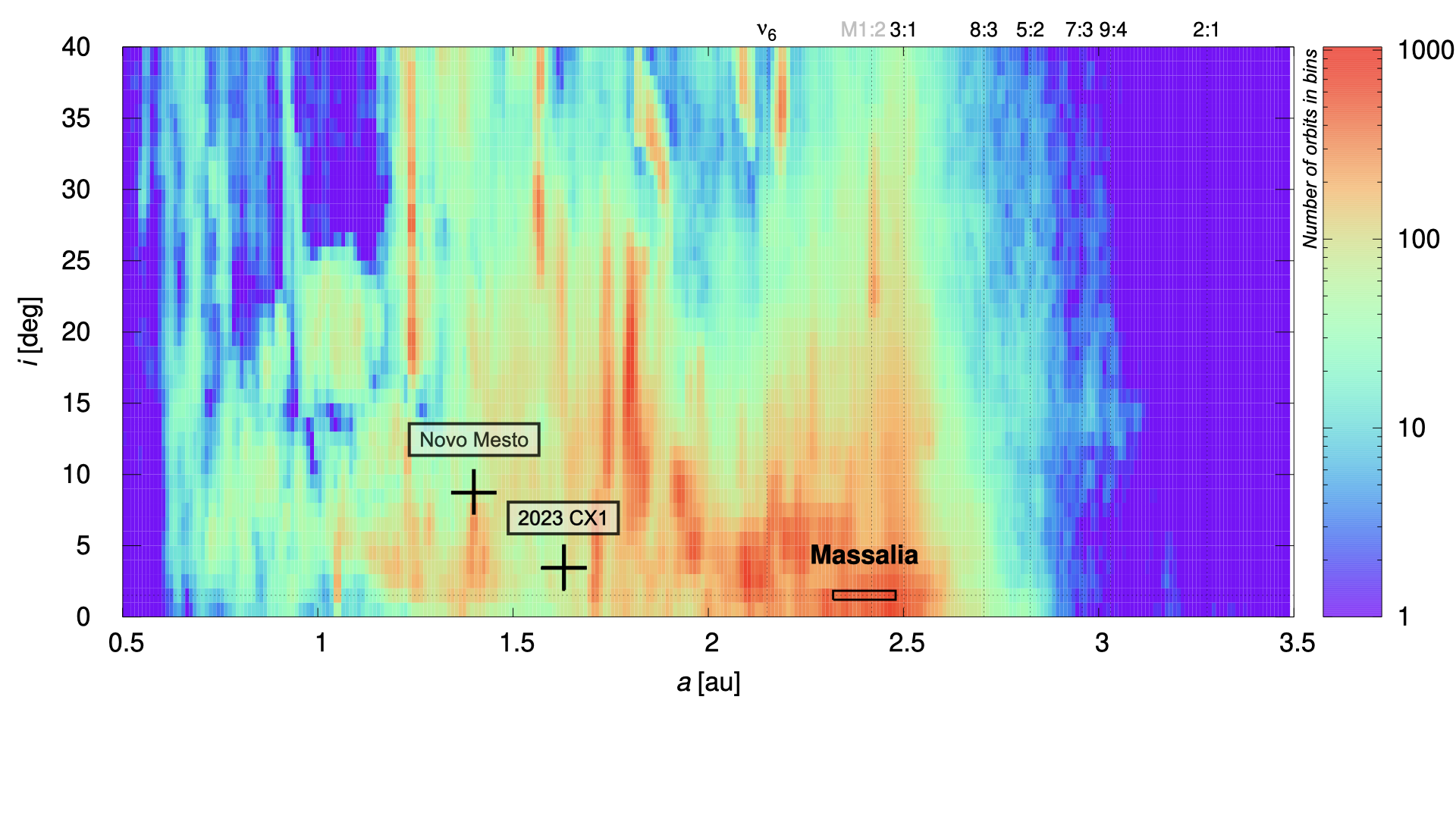} \\
\includegraphics[trim={0.8cm 5.5cm 0cm 1cm},clip,width=.9\textwidth]{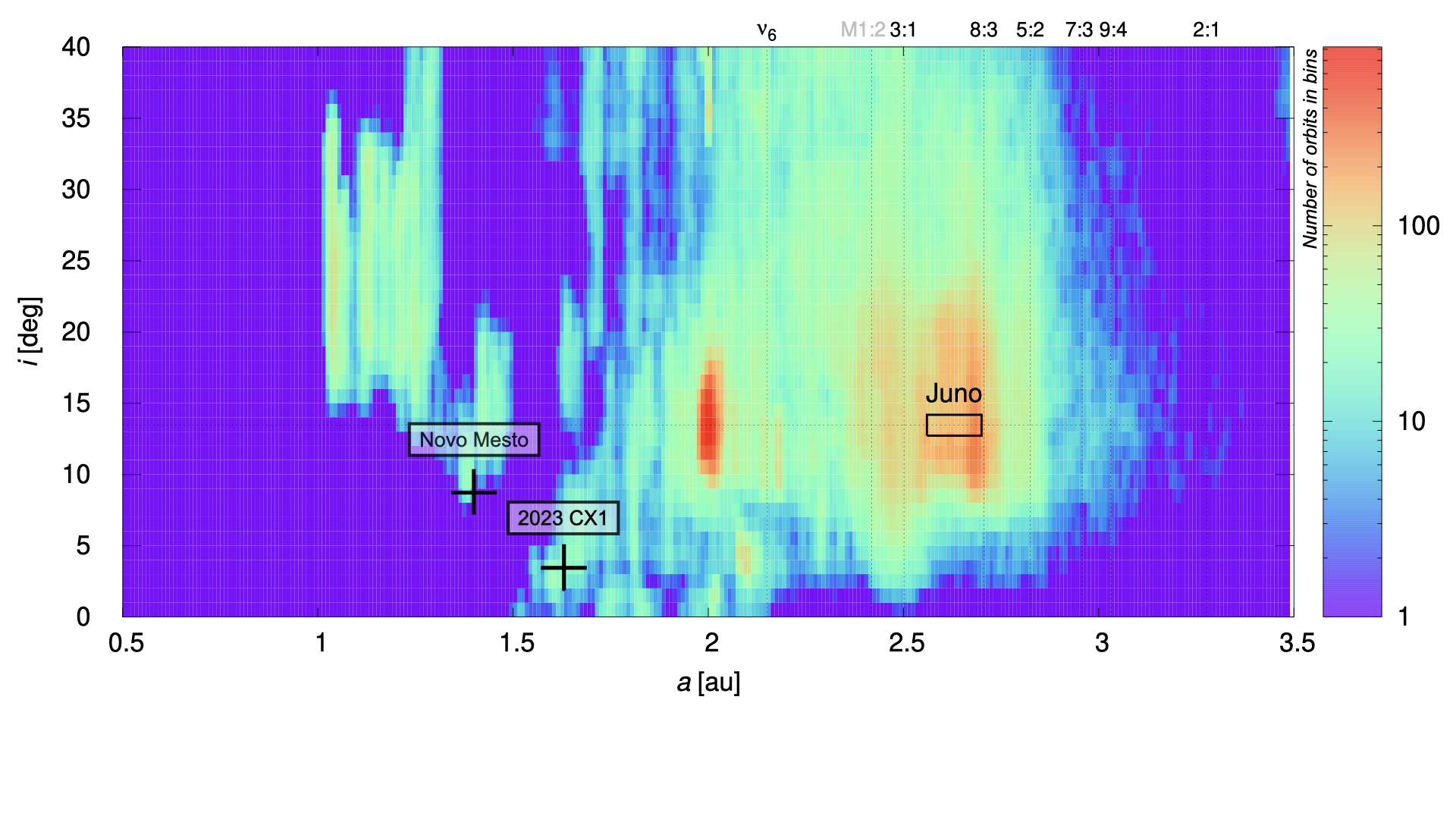} \\
\includegraphics[trim={0.8cm 5.5cm 0cm 1cm},clip,width=.9\textwidth]{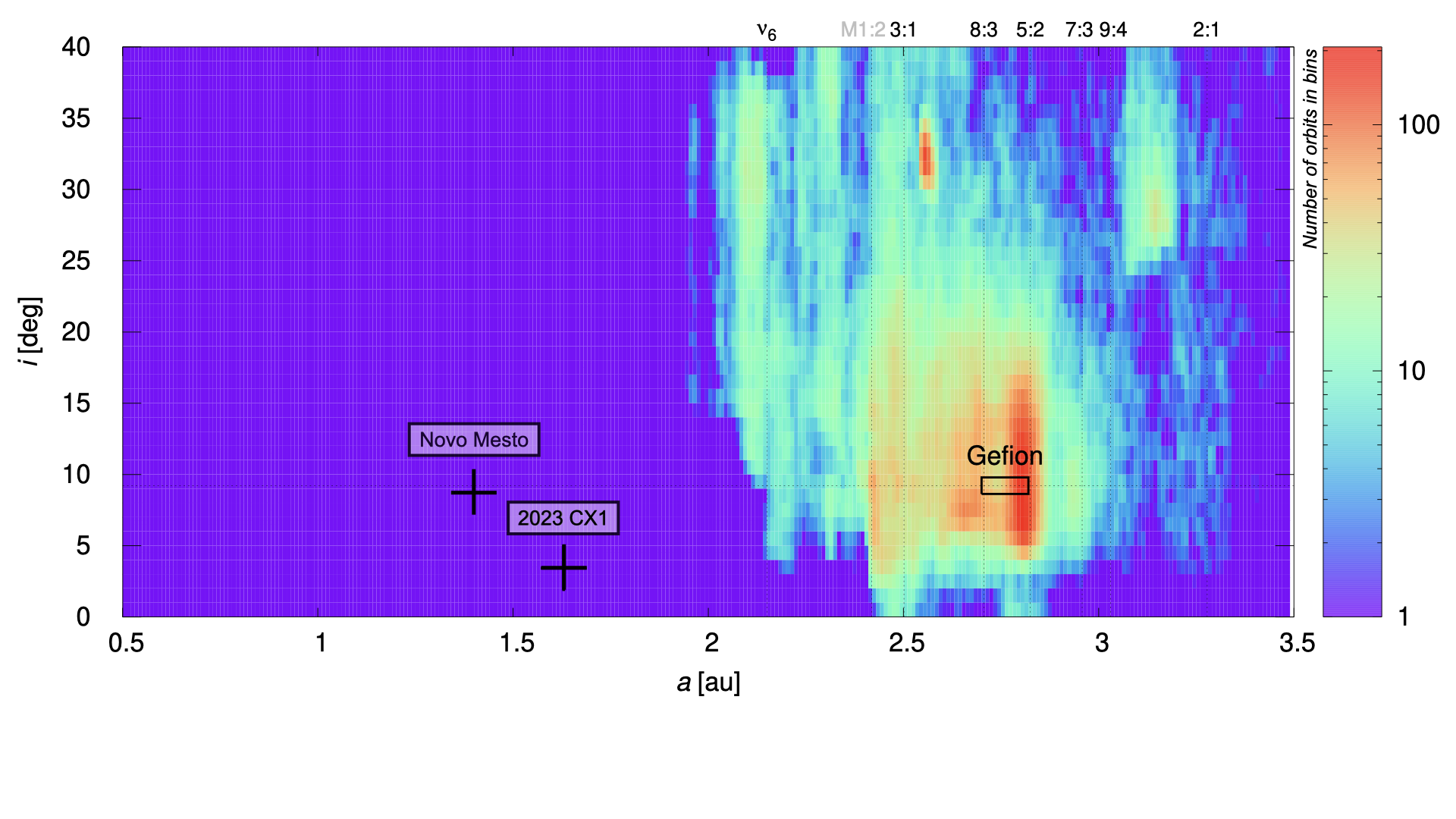} \\
\end{tabular}
\caption{Synthetic probability maps for an orbit with a given semi-major axis~$a$ and inclination~$i$ to originate from one of the L chondrite-like asteroid families: Massalia$2$ (top), Juno (middle), and Gefion (bottom). The precise probability $p$ of originating from a source can be computed by normalizing and multiplying each map by the corresponding population of $N{\rm neo}({>}1{\rm m})$ bodies in the NEO region. Results for the Saint-Pierre-Le-Viger and Novo Mesto meteorites, computed using the METEOMOD software (\url{https://sirrah.troja.mff.cuni.cz/~mira/meteomod/meteomod.html}), are provided in Table \ref{tab:meteomod}. The Massalia$_2$ family produces low-inclination orbits compatible with the preatmospheric trajectories of the Saint-Pierre-Le-Viger and Novo Mesto meteorites.}
\label{fig:meteomod_map}
\end{figure}

\begin{figure}[ht]
     \centering
     \includegraphics[width=.99\textwidth]{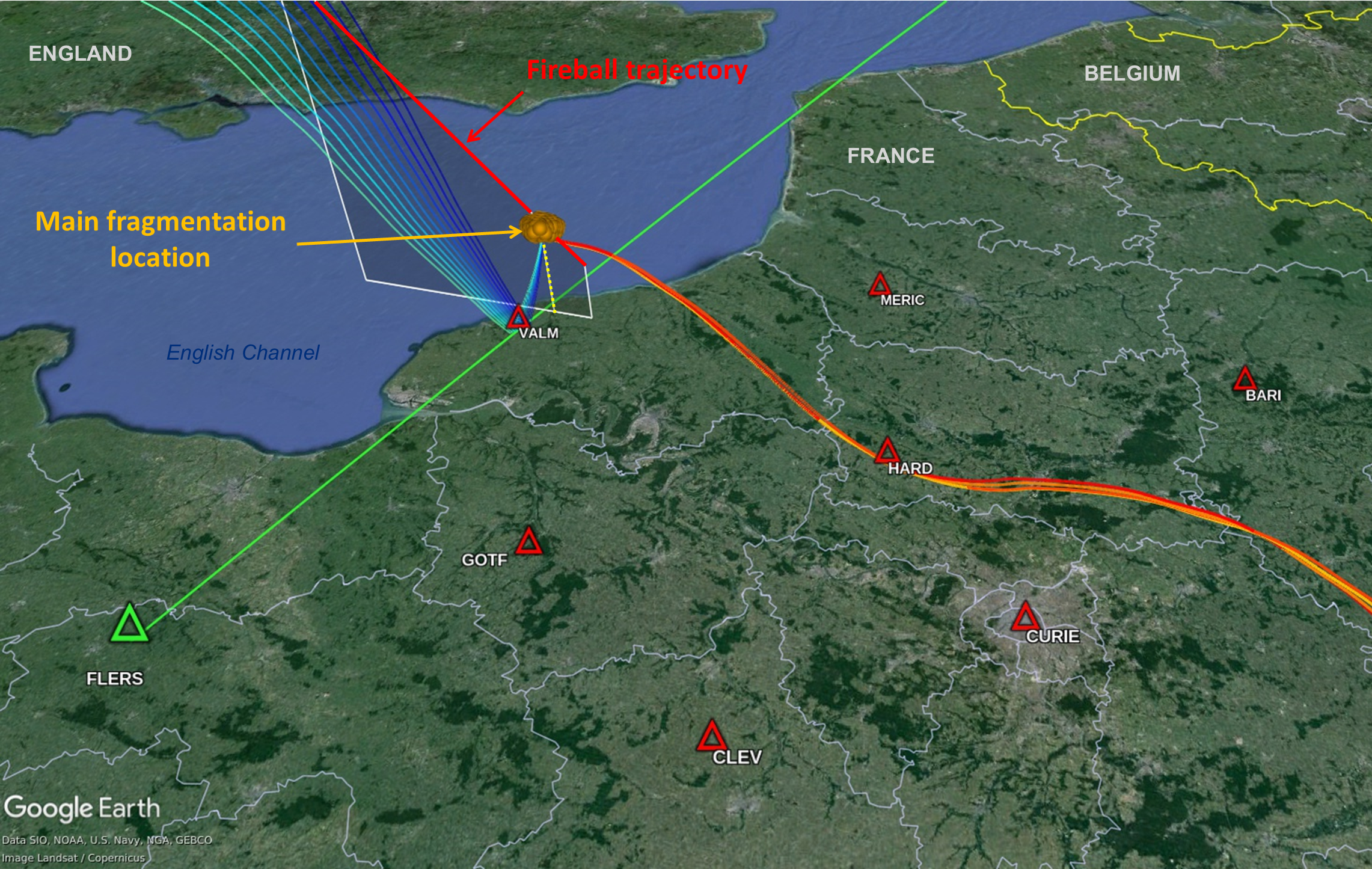}
     \caption{3D acoustic location results of main fragmentation using realistic station-dependent 3D-celerity models from arrivals picked on the eight stations depicted as triangles. Red triangles are seismic accelerometers of the RESIF French permanent broadband seismic network, green triangle is the FLERS French infrasound experimental array. Orange points on the bolide trajectory are solutions obtained from different 3D-celerity models corresponding to time residuals less than 8 seconds. Acoustic rays are examples of eigen rays calculated for stations VALM and HARD which were identified to calculate 3D-celerity models. Altitude of best location is found at 27.9$\pm$1.0 km.}
     \label{fig:3D_acoustic}
 \end{figure}

Additionally, we used the METEOMOD orbital distribution model for meteoroids \cite{Broz2024} to estimate the probability that 2023 CX1 originated from one of the asteroid families located between 1.9 and 3.5 au. As detailed in \cite{Marsset2024}, probability distribution maps in Near-Earth Object (NEO) space (semi-major axis and inclination) were employed to identify the source region of L chondrite orbits (Table \ref{tab:meteomod} and Figure \ref{fig:meteomod_map}). This method differs from those used in previous models (e.g. \cite{Granvik2018,Nesvorny2023}, where sources correspond to resonances (e.g., $\nu_6$, 3:1, 5:2, 2:1), thereby losing much of the information about the original orbital elements ($a$, $e$, $i$). In addition, METEOMOD utilizes the size-frequency distributions (SFDs) of the sources, which serve as weights, and incorporate the taxonomy of the sources to prevent ambiguities.
The probability of originating from a selection of S-type asteroid families, including the recent subpopulations of the Massalia and Koronis families (Massalia$_2$ and Koronis$_2$) as identified in \cite{Broz2024}, is provided in Supplementary Information (Table \ref{tab:meteomod}). 

\subsection{Blastwave pressure calculations} \label{sec:airburst_pressure}

The blast wave propagation to the ground was done using the ALE3D hydrocode on the Aitken Supercomputer at NASA Ames. Energy was deposited according to the meteor lightcurve profiles for an adopted entry speed of 14 km/s and a total energy of 18 Mt. A constant luminous efficiency was assumed, so the energy deposition rate dE/dt was proportional to the luminosity. A domain was filled with ideal gas air using an isothermal atmosphere at 255 K and a scale height of 7.5 km. Simulation cells were aligned with the trajectory at 49.1$\degree$,  and the energy deposited uniformly during the time the meteor was passing through each cell.  For a larger meteor there may also be a significant amount of momentum imparted, but that was neglected for these simulations. 

To efficiently propagate the blast wave to the ground, the energy deposition was scaled up to 18.4 Mt, producing a blast wave kilometers in length and allowing a much coarser mesh. While this energy represents approximately one million times the kinetic energy of 2023 CX1, the distribution of overpressure on the ground is expected to be similar to that of 2023 CX1, although at  lower pressures. A real 18 Mt impactor would burst lower in the atmosphere than the energy profile simulated here. At the actual energy of 2023 CX1, the resulting shock waves would have lower overpressure and be narrower, but would still originate from the same points along the trajectory. Pressure waves travel faster at higher overpressure, but even in the 18 Mt case they drop to near sonic speed within a few kilometers of the source. This means that the pressure waves from a 2023 CX1-like event will combine in a similar pattern to those of the 18 Mt case, resulting in a comparable ground footprint, but at non-damaging overpressure levels likely within background noise.

\backmatter

\backmatter

\section*{Data Availability Statement}

Updated astrometric measurements of asteroid 2023 CX1 prior to impact are publicly available at the Minor Planet Center (available at \url{https://minorplanetcenter.net}). The updated orbital solution is accessible via the JPL Small-Body Database Browser at \url{https://ssd.jpl.nasa.gov/tools/sbdb_lookup.html#/?sstr=2023%20CX1}. Precise coordinates and masses of the retrieved meteorites, as well as all relevant laboratory measurements, are provided as Extended Data or Supplementary Information files. Data supporting the fireball trajectory, photometry, strewn field computations and seismo-acoustic analysis are available on Zenodo \cite{zenodo2025}  \color{blue}{https://doi.org/10.5281/zenodo.15328378}\color{black}. Additional data sets generated during and/or analyzed during the current study are available from the corresponding author upon reasonable request. 

\section*{Code Availability Statement}

The proprietary software used in this analysis includes methods for computing the asteroid's orbit, analyzing the fireball's fragmentation, modeling the strewn field, and simulating the blast wave propagation using ALE3D (\url{https://sd.llnl.gov/stockpile-science/high-performance-computing/proprietary-software}). Calibration of part of the fireball optical data was performed using the open-source \texttt{SkyFit2} software, which is part of the \texttt{RMS} library available at \url{https://github.com/CroatianMeteorNetwork/RMS}. Independent confirmation of the fireball trajectory parameters was performed using the \texttt{WesternMeteorPyLib} (\texttt{wmpl}), which was also used to compute the trajectory and support the strewn field modeling. It is available at \url{https://github.com/wmpg/WesternMeteorPyLib}. Probabilities of originating from the Massalia asteroid family were computed using the METEOMOD model, accessible at \url{https://sirrah.troja.mff.cuni.cz/~mira/meteomod/meteomod.php}. \\

Supplementary Information is available for this paper. Correspondence and requests for materials should be addressed to Auriane Egal.

\bmhead{Acknowledgments} 

We thank M. Alexandersen (MPC) for reviewing and correcting station coordinates in the MPC database. Model wind data was provided by R. Bro\v{z}kov\'a from the Czech Hydrometeorological Institute. Data visualization was provided by P. Geoffroy. E. J. Christensen, R. L. Seaman, and J. Stone for providing observations from the Catalina Sky Survey. The following camera operators contributed 2023 CX1 imaging to the FRIPON, GMN and UKMON networks: M. Rushton, A. Pratt, S. Saunders, J. Olver, N. Russel, and M. McIntyre. Targeted fireball observation were made by V. Devillepoix, J-L. Devillepoix, P. Devillepoix, A. Favre, A. Greenway, L. Greenway, Y. Trotel and P. Wright, amongst others. Observatori del Montsec-IEEC recorded the fireball from Sant Esteve de la Farga, Lleida, Catalonia at a nominal distance of 900 km. B. Gladman for dynamical computations of the asteroid's escape route. F. Nikodem, L. Smuła and M. Smuła for providing the coordinates and masses of the fragments found. R. Wieler for contributing to the discussion of the noble gas results. The help of the VERA “cosmogenics” team, especially Stephanie Adler, Robin Golser, Peter Steier, and Carlos Vivo-Vilches is highly appreciated. The meteorite field search was supported by members of the FRIPON/Vigie-Ciel citizen science network, with participants listed at \url{https://www.fripon.org/wp-content/uploads/2025/09/2023CX1-contributors.pdf}.

AEg, DVi, PBr, PWi were supported in part by the NASA Meteoroid Environment Office under cooperative agreement 80NSSC21M0073. JBo and PSp were supported by grant No.\ 19-26232X from the Czech Science Foundation. SAn received funding from the European Union’s Horizon Europe research and innovation program under the Marie Skodowska-Curie grant agreement No. 101150536, for the project 'FLAME'. DFa and SNa conducted this research at the Jet Propulsion Laboratory, California Institute of Technology, under a contract with the National Aeronautics and Space Administration (80NM0018D0004). The Czech Science Foundation supported this research through grants 25-16507S (MBr) and 25-16789S (JHa). Additional funding was received from the Hungarian National Research, Development, and Innovation Office grant K-138962 and GINOP-2.3.2-15-2016-00003, and from the Slovak Science and Grant Agency, project VEGA-1/0487/23 (for PPo, ISy and IKo). DK, LME and HB were supported by the Swiss SNF and NCCR PlanetS under grants 51NF40\_205606 and SNF\_219860. The VERA \quotes{cosmogenics} team was supported by the ChETEC-INFRA (Horizon 2020, No. 101008324) grant. TSR acknowledges funding from Ministerio de Ciencia e Innovación (Spanish Government), PGC2021, PID2021-125883NB-C21. This work was (partially) supported by the Spanish MICIN/AEI/10.13039/501100011033 and by \quotes{ERDF A way of making Europe} by the \quotes{European Union} through grant PID2021-122842OB-C21, and the Institute of Cosmos Sciences University of Barcelona (ICCUB, Unidad de Excelencia \quotes{María de Maeztu}) through grant CEX2019-000918-M. JMTR acknowledges support from the Spanish project PID2021-128062NB-I00 funded by MCIN/AEI. A.J. acknowledges support from ANID Millennium Science  Initiative ICN12\_009, IM23-0001. RAM acknowledge support from FONDECYT/ANID grant \# 124 0049 and from ANID, Fondo GEMINI, Astrónomo de Soporte GEMINI-ANID grant \# 3223 AS0002. DHu and JJa were supported in part by the SIAA \quotes{Foundation Interactive Astronomy and Astrophysics} in Tübingen, Germany. The Joan Oró Telescope (TJO) at the Montsec Observatory (OdM) is owned by the Catalan Government and operated by the Institute of Space Studies of Catalonia (IEEC). The UK Fireball Alliance meteor camera observatory was supported by the Science and Technology Facilities Council [grant number:ST/Y004817/1]. PJe was supported by NASA grant 80NSSC18K0854. \\

This preprint has not undergone peer review (when applicable) or any post-submission improvements or corrections. The Version of Record of this article is published in \textit{Nature Astronomy}, and is available online at https://doi.org/10.1038/s41550-025-02659-8.

\section{Author contributions}

A. Egal, D. Vida, F. Colas, B. Zanda coordinated the research. A. Egal, D. Vida, B. Zanda and P. Jenniskens drafted the paper. M. Micheli, K. S\'arneczky, A. P\'al, D. Farnocchia, S. Naidu, L. Conversi, F. Ocaña, D. Husar, J. Jahn, P. Dupouy, T. Santana-Ros and A. Egal collected and calibrated the data associated with asteroid discovery, imaging and astrometry. N. Moskovitz, T. Kareta, T. Santana-Ros, J. Hanuš, M. Devogèle, L. Buzzi, K. Korlevi\'{c}, M. Birlan, D.A. Nedelcu, A. Sonka and F. Losse collected and calibrated the data related to the asteroid photometry. P. Wiegert, P. Vernazza, M. Marsset, M. Brož, P. Shober, A. Lagain and A. Egal completed the asteroid dynamical analysis. J. Borovi\v{c}ka, P. Spurný, D. Vida, H.A.R. Devillepoix, A. Egal, S. Anghel, R. Neubert, M. McIntyre, D. Šegon, J. Vaubaillon, K. Bailli\'{e}, J. Desmars and the FRIPON International Team (FIT) collected and analysed the data relevant to meteor triangulation and photometry. J. Borovi\v{c}ka, P. Spurný, D. Vida and H.A.R. Devillepoix undertook the fragmentation and strewn-field calculation. F. Colas, S. Bouley, A. Steinhausser, B. Zanda, P. Vernazza, J. Vaubaillon, A. Malgoyre, K. Antier, L. Maquet, J. Desmars, K. Bailli\'{e}, M. Birlan, S. Bouquillon, S. Jeanne, P. Beck, P. Jenniskens and FIT collected and calibrated the data associated with meteorite recovery and search campaign coordination. P. Brown, J. Assink and L. Evers collected and calibrated the data for the infrasound analysis. A. Le Pichon, G. Mazet-Roux, J. Vergoz, J. Vergne and L. McFadden collected and calibrated the data related to the seismo-acoustic analysis. B. Zanda, J. Gattacceca, L. Ferrière, M. Gounelle, S. Pont, I. Baziotis, J.-A. Barrat and P. Sans-Jofre collected and calibrated the data relevant to meteorite sampling and characterization. D. Krietsch, H. Busemann, C. Maden and L.M. Eckart collected and calibrated the data in the analysis of noble gas isotopes in the meteorites. P. Povinec, I. S\'ykora, I. Kontul', O. Marchhart, M. Martschini, S. Merchel and A. Wieser collected and calibrated the data in the analysis of cosmogenic nuclides in the meteorites. S. de Vet collected and calibrated the data in the analysis of meteorite density. D. Robertson undertook the blast-wave pressure calculations. P. Jenniskens, O. Hernandez and P. Brown provided general guidance and improvements to the paper. All authors reviewed the results and approved the final version of the paper. \\

\textbf{FRIPON International Team} \\[0.1cm]
 Josep Maria Trigo-Rodriguez$^{1,2}$, Enrique Herrero$^{2}$, Jim Rowe$^{3}$, Andrew R.D. Smedley$^{3,4}$, Ashley King$^{3,5}$, Salma Sylla$^{6}$, Daniele Gardiol$^{7}$, Dario Barghini$^{7}$, Hervé Lamy$^{8}$, Emmanuel Jehin$^{9}$, Detlef Koschny$^{10}$, Bjorn Poppe$^{11}$,  Andrés Jordán$^{12,13,14}$, Rene A. Mendez$^{15}$, Katherine Vieira$^{16}$, Hebe Cremades$^{17}$, Hasnaa Chennaoui Aoudjehane$^{18}$, Zouhair Benkhaldoun$^{19}$.\\

\small{$^1$Institute of Space Sciences (CSIC), Campus UAB, Carrer de Can  Magrans, s/n 08193 Cerdanyola del Vallés (Barcelona), Catalonia, Spain $^2$Institut d’Estudis Espacials de Catalunya (IEEC), Esteve Terradas 1, Edificio RDIT, Ofic. 212 Parc Mediterrani de la Tecnologia (PMT) Campus del Baix  Llobregat – UPC 08860 Castelldefels (Barcelona), Catalonia, Spain $^{3}$The UK Fireball Alliance (UKFAll) $^{4}$Department of Earth and Environmental Sciences, University of Manchester, Manchester, M13 9PL UK $^{5}$ Natural History Museum, London, SW7 5BD, UK $^{6}$Université Cheikh Anta Diop Dakar-Fann, Sénégal $^{7}$INAF, Osservatorio Astrofisico di Torino, Via Osservatorio 20, Pino Torinese, 10025, Italy $^{8}$Royal Belgian Institute of Space Aeronomy, Uccle, Belgium $^{9}$STAR, Space sciences, Technologies Astrophysics Research Institute, Université de Liège, Liège, 4000, Belgium $^{10}$Lunar and Planetary Exploration, Technical University of Munich, Lise-Meitner-Str. 9, D-85521 Ottobrunn, Germany $^{11}$Division for Medical Radiation Physics and Space Environment, University of Oldenburg, Oldenburg, Germany $^{12}$Facultad de Ingenier\'ia y Ciencias, Universidad Adolfo Ib\'{a}\~{n}ez, Av. Diagonal las Torres 2640, Pe\~{n}alol\'{e}n, Santiago, Chile $^{13}$Millennium Institute for Astrophysics, Santiago, Chile $^{14}$El Sauce Observatory --- Obstech, Coquimbo, Chile $^{15}$ Departamento de Astronomía, Universidad de Chile, Casilla 36-D, Santiago, Chile $^{16}$Instituto de Astronomía y Ciencias Planetarias, Universidad de Atacama, Copayapu 485, Copiapó, 1531772, Chile $^{17}$CONICET, Grupo de Estudios en Heliofisica de Mendoza, Universidad de Mendoza, Boulogne Sur Mer 665, Mendoza, 5500, Argentina $^{18}$Hassan II University of Casablanca, GAIA Laboratory, Faculty of Science Ain Chock, km 8 route d’El Jadida, 20150 Casablanca, Morocco  $^{19}$Oukaimeden Observatory, High Energy Physics and Astrophysics Laboratory, Faculty of Science Semlalia,  Cadi Ayyad University, Marrakech, Morocco. }

\section*{Declaration of Competing Interests}

The authors declare no competing interests.\\

\newpage
\newpage
\section{Extended data}

\begin{table}[h]
\caption{Estimates of the preatmospheric mass, radius, and geometric albedo ($\rho_V$) of 2023 CX1 derived with multiple independent methods, as detailed in Section \ref{sec:preatm_size}. Within uncertainties, most estimates converge toward a preatmospheric radius between 35 and 39 cm. The only exception is the estimate based on measured $^{60}$Co activity via gamma spectrometry (GS), which yields a lower value, likely due to the meteoroid being too small to develop a full nucleonic cascade in its interior.  A weighted average combining the estimates from asteroid photometry, fireball light curve, infrasound data, and $^{26}$Al and $^{60}$Co GS yields a preferred preatmospheric radius of 36 $\pm$ 3 cm and an initial mass of 650 $\pm$ 160 kg. \color{black}}\label{tab:mass}%
\begin{center}
\begin{tabular}{@{}llllllll@{}}
\toprule
 Technique / Data  & Mass (kg) & Radius (cm) & Albedo $p_V$\footnotemark[1] \\
\midrule
Asteroid magnitude & 1280 $\pm$ 780 & 43 $\pm$ 10   & (0.196 ± 0.036)\footnotemark[2]\\[0.1cm]
Infrasound & 1450 $\pm$ 850 & 45 $\pm$ 10 & 0.18 ± 0.09\\[0.1cm]
Lightcurve &  650 $\pm$ 160 & 36 $\pm$ 3  & 0.28 ± 0.05\\[0.1cm]
Cosmogenic $^{21}$Ne/$^{22}$Ne & $\ge$ 370 & $\ge$ 30  & $\le$ 0.41\\[0.1cm]
$^{26}$Al activity (IAMS) & $\ge$ 110 &  $\ge$ 20 & $\le$ 0.92\\[0.1cm]
$^{26}$Al activity (GS) &  470 $\pm$ 410 & 28 $\pm$ 12   & 0.47 ± 0.60\\[0.1cm]
$^{60}$Co activity (GS) & 230 $\pm$ 100 & 25 $\pm$ 4 & 0.59 ± 0.20\\
\botrule
\end{tabular}
\end{center}
\footnotetext[1]{Size to geometric albedo conversion was performed following \cite{Bowell1989}, for a measured absolute magnitude H of 32.7$\pm$0.3. Uncertainties were symmetrized by averaging the propagated upper and lower bounds. Calculations of radius assume a spherical body with a density of 3300 kg/m$^3$. }
\footnotetext[2]{Assumed geometric albedo range for 2023 CX1, based on the preferred albedo for asteroid (20) Massalia reported by the the Virtual Observatory Solar System Open Database Network \cite{Berthier2023}.}
\end{table}

\begin{table}[ht]
\caption{Atmospheric trajectory of the fireball before and after the flare. The coordinates, including altitude, are given in the WGS84 system. The slope is given relative to the local horizon. The velocity is given relative to the Earth surface. The brightest part of the fireball was excluded from the measurement.}\label{tab:trajectory}
\begin{tabular}{llrr}
\toprule
  \multicolumn{2}{c}{\textit{Before the flare}} & Beginning & End  \\
  \cmidrule{3-4}
  Latitude & ($\degree$N) & 49.92182 $\pm ~0.00015$ & 49.80774 $\pm ~0.00015$  \\ 
  Longitude & ($\degree$E) & $-$0.16713 $\pm ~0.00015$ & 0.68323 $\pm ~0.00015$ \\
  Altitude & (km) & 101.755 $\pm ~0.010$ & 29.412 $\pm ~0.010$  \\
  Azimuth & ($\degree$) & 281.400 $\pm ~0.015$  & 282.050 $\pm ~0.015$ \\
  Slope & ($\degree$) & 49.098 $\pm ~0.010$ & 48.725 $\pm ~0.010$ \\
  Velocity & (km/s) & 14.04 $\pm ~0.03$ & 13.36 $\pm ~0.20$\\
  \multicolumn{2}{c}{\textit{After the flare}} &  &   \\[-0.2cm]
  \cmidrule{3-4}
  Latitude & ($\degree$N)  & 49.80306 $\pm~0.00025$& 49.78903 $\pm~0.00020$ \\ 
  Longitude & ($\degree$E) & 0.71441  $\pm~0.00015$ & 0.80614 $\pm~0.00015$\\
  Altitude & (km) & 26.760 $\pm~0.020$ & 18.9119 $\pm~0.015$ \\
  Azimuth & ($\degree$) & 283.27 $\pm~0.12$ & 283.34 $\pm~0.12$ \\
  Slope & ($\degree$) & 49.04 $\pm~0.06$ & 49.21 $\pm~0.10$ \\
  Velocity & (km/s) & 10.25 $\pm~0.25$ & 2.68 $\pm~0.20$ \\
 \midrule
 Apparent Right Ascension  & ($\degree$) &   $128.286 \pm  0.015$\\
Apparent Declination & ($\degree$) & $41.505 \pm  0.015$    \\ 
Initial Velocity & (km/s) & $14.04 \pm 0.03$ \\
Geocentric Right Ascension & ($\degree$)  &   $117.25 \pm  0.07$\\
Geocentric Declination & ($\degree$) & $34.81 \pm  0.05$    \\  
Geocentric Velocity & (km/s) & $8.91 \pm 0.05$ \\
 \botrule
\end{tabular}
\end{table}

\begin{table}[ht]
\caption{Fragmentation details of 2023 CX1. The first three rows compare the timing and locations of fragmentation events identified from optical observations of the fireball, along with an independent determination of the main fragmentation event derived from acoustic measurements (see Section \ref{sec:seismo-acoustic_methods}). The acoustic estimate shows good agreement with the optical results within uncertainties. The remaining rows present the separation heights and initial masses of the nine fragments identified in the video recorded from the village of {\it La Fresnaye}, along with the masses of meteorites recovered near their predicted landing sites.}
\label{tab:fireball_fragmentation}
  \resizebox{\textwidth}{!}{\begin{tabular}{lllll}
  \cmidrule{2-5}
  \textit{Fragmentation location} & Time (UT)  & Longitude ($\degree$) &  Latitude ($\degree$) & Height (km) \\
  \midrule
  Optical (1$^{st}$ fragmentation) & 02:59:20.10 $\pm$ 0.04 s & 0.683 $\pm$ 0.00015 & 49.808 $\pm$ 0.00015 & 29.4 $\pm$ 0.01\\
  Optical (2$^{nd}$ fragmentation) & 02:59:20.23 $\pm$ 0.04 s & 0.699 $\pm$ 0.00015 & 49.806 $\pm$ 0.00025 & 28.1 $\pm$ 0.02\\
  Acoustic & 02:59:17.4 $\pm$ 4 s & 0.64 $\pm$ 0.07 & 49.81 $\pm$ 0.05 & 27.9 $\pm$ 1.0 \\
  \cmidrule{1-5}
  \textit{Individual fragments} & Label & Origin height (km) & Initial mass (g) & Meteorite mass (g) \\

  & A & 29.4 & 5500 & 1550 \\
  & B & 28.1 & 960 & 630 \\
  & C & 28.1 & 700 & 460 \\
  & D & 29.4 & 620 & 400 \\
  & E & 28.1 & 610 & 400 \\
  & F & 34.6 & 300 & 190 \\
  & G & 28.1 & 190 & 130 \\
  & H & 28.1 & 90 & 60 \\
  & I & 29.4 & 70 & 50 \\

 \botrule
 \end{tabular}
 }
 \end{table}

\begin{table}[!ht]
 \caption{Summary of infrasound signal characteristics from the 2023 CX1 fireball. Measurements from the Raspberry Pi Shake \& Boom citizen program have preamble \quotes{AM-}.}
 \label{tab:infrasound} 
\resizebox{\textwidth}{!}{\begin{tabular}{@{}lllllllll@{}}
  \toprule
  Station   & Range & Max. Amp. & Az Dev. & Peak-to-peak &  Period at Max Amp & Period at  & Celerity \\
            &  (km) & (Pa) & (degs) & amplitude (Pa)  & (zero-crossing)(sec) & peak PSD (sec) & (km/s)\\
  \midrule
  I48TN    &1701 &0.04 $\pm$ 0.015 &1.5 &0.06 $\pm$ 0.03 &1.2 $\pm$ 0.3 &1.5 &0.316 &  \\
  I26DE    &941 &0.25 $\pm$ 0.18 &-1 &0.43 $\pm$ 0.36 &1.64 $\pm$ 0.03 &1.8 &0.321 &  \\
  I43RU    &2507 &0.05 $\pm$ 0.01 &-3.2 &0.08 $\pm$ 0.02 &1.61 $\pm$ 0.01 &1.5 &0.293 &  \\
  I31KZ    &3969 &0.006 $\pm$ 0.005 &-23.1 &0.011 $\pm$ 0.01 &2.18 $\pm$ 0.12 &1.3 &0.292 &  \\
  I46RU    &5439 &0.008 $\pm$ 0.006&-3.2&0.04 $\pm$ 0.012 &1.83 $\pm$ 0.62 &1.6 &0.291 &  \\
  DBNI     & 450 & 0.29 $\pm$ 0.02 & - & 0.42 $\pm$ 0.03 & 0.67 $\pm$ 0.03 & 0.71  & 0.292 &  \\
  FLERS    & 145 & 0.40 & - & 0.76 & 1.6 & 1.5 & - &  \\
  CEA      & 170 & 2.67 & - & 4.24 & 1.4 & 1.37 & - &  \\
  AM-R9BD8 & 165 & 0.12 & - & 0.577 & 0.35 & 0.27 & 0.310 &  \\
  AM-R82EC & 180 & 0.27 & - & 0.509 & 0.72 & 0.79 & 0.301&  \\
  AM-RD5DB & 165 & 0.34 & - & 0.616 & 0.37 & 0.36 & 0.296 &  \\
  AM-RD44E & 160 & 0.32 & - & 0.954 & 0.25 & 0.34 & 0.314 &  \\
\botrule
 \end{tabular}
   }
\end{table}

\begin{table}[!ht]
\caption{Arrival features of the seismo-acoustic signal detected by several stations in France and United Kingdom. Only stations used to locate the main fragmentation event are presented in the table. Listed are range (R), Peak-to-peak amplitudes (Amp.), which are read on the vertical component, Back-Azimuth (Az.), 3D Celerity model distance, corresponding horizontal distance, wave velocity, model wave velocity, calculated residual in back azimuth.}
\label{tab:seismo_acoustic} 
\resizebox{\textwidth}{!}{\begin{tabular}{cccccccccc}
  \toprule
Station	 & 	Time	& R & Amp. & Az. & Dist  & 	HDist  & 	V & Vc & $\Delta$Az  \\
        &        & 	(km) & (nm/s) &  (°) &  (km) & 	 (km)  &    (km/s) & 	  (km/s) & degs \\
\midrule
VALM	  & 	03:01:05	& 28 & 1236 & 252 & 30.9	& 	11.4  	 & 0.280	 & 0.281 & 	-0.3	 \\
GOTF	  & 	03:04:23	 & 89 & 117 & 181 & 93.4	& 	88.9  	 & 0.299	 & 0.302	 & -3.2	 \\
MERIC	  & 	03:04:36	& 89 & 310 & 82 & 100.3	& 	96.1  	 & 0.315	 & 0.311	 & 3.8	 \\
HARD	  & 	03:05:11	& 103 & 310 & 123 & 114.5	& 	110.9 	 & 0.313	 & 0.320	 & -8.1  \\
FLERS	  & 	03:07:13	& 145 & - & 33.2 & 144.1	& 	141.2  & 	0.304		 & 0.302	&  3.1 \\
CLEV	  & 	03:07:29	& 147 & 413 & 161 & 153.7	& 	151.0 	 & 0.310 & 	0.310 & 	-0.5 \\
CURIE	  & 	03:08:14 &  - & 	 - & -		& 167.7	& 	165.3  & 0.317	 & 0.310	 & 12.1 \\
BARI		 &    03:09:12	& 188 & 1006 & 97 & 198.7	&   196.7 	 & 0.330 & 	0.332 & 	-4.4  \\
\botrule
   \end{tabular}}

\end{table}

\begin{table}[!ht]
\caption{
 He, Ne, Ar, Kr, Xe concentrations and isotopic ratios. He, Ne, Ar concentrations are given in 10$^{-8}$ cm$^3$ STP/g, Kr and Xe concentrations are provided in 10$^{-10}$ cm$^3$ STP/g. Shielding-dependent measured cosmogenic isotope ratios are also provided.}
\label{tab:noble_gas_concentrations}
\resizebox{\textwidth}{!}{\begin{tabular}{@{}lrrrrr@{}}
\toprule
SPLV sample & \#4 & \#10 & \#13 & \#26 & \#1 \\
\midrule
Sample mass [mg]	& 20.157$\pm$0.008 & 16.647$\pm$0.013 & 17.763$\pm$0.015 & 23.153$\pm$0.018 & 16.061$\pm$0.011 \\
\midrule
$^{4}$He						& 877.3$\pm$2.4  & 1228.3$\pm$3.6 & 1251.0$\pm$4.1 & 1156.0$\pm$3.8 & 1100.3$\pm$3.1 \\
$^{3}$He/$^{4}$He$\times10^{4}$	& 613.9$\pm$2.4 & 433.7$\pm$1.9 & 404.4$\pm$1.9 & 441.5$\pm$2.0 & 469.5$\pm$2.1 \\
$^{20}$Ne						& 11.560$\pm$0.061 & 11.799$\pm$0.067 & 11.620$\pm$0.077 & 11.550$\pm$0.057 & 10.806$\pm$0.070 \\ 
$^{20}$Ne/$^{22}$Ne 			& 0.8392$\pm$0.0039	& 0.8334$\pm$0.0039 & 0.8360$\pm$0.0044	 & 0.8461$\pm$0.0033 & 0.8374$\pm$0.0040 \\
$^{21}$Ne/$^{22}$Ne 			& 0.9336$\pm$0.0039 & 0.9335$\pm$0.0042 & 0.9263$\pm$0.0044 & 0.9332$\pm$0.0039 & 0.9282$\pm$0.0046 \\
($^{22}$Ne/$^{21}$Ne)$_{cos}$	 			& 1.0711$\pm$0.0044 & 1.0713$\pm$0.0048	& 1.0795$\pm$0.0051 & 1.0716$\pm$0.0045 & 1.0774$\pm$0.0053 \\
($^{3}$He/$^{21}$Ne)$_{cos}$ 	& 4.188$\pm$0.023 & 4.031$\pm$0.026 & 3.929$\pm$0.028 & 4.007$\pm$0.024 & 4.313$\pm$0.032 \\
\midrule
$^{36}$Ar 						& 1.573$\pm$0.017 & 2.175$\pm$0.021 & 2.503$\pm$0.029 & 1.638$\pm$0.017 & 2.390$\pm$0.021 \\
$^{36}$Ar/$^{38}$Ar 			& 0.978$\pm$0.011 & 1.148$\pm$0.011 & 1.426$\pm$0.017 & 0.970$\pm$0.010 & 1.525$\pm$0.014 \\
$^{40}$Ar/$^{36}$Ar 			& 4110$\pm$55 & 2854$\pm$34 & 2649$\pm$38 & 3921$\pm$50 & 2661$\pm$31 \\
$^{84}$Kr 						& 0.789$\pm$0.019 & 0.578$\pm$0.022 & 1.007$\pm$0.021 & 0.648$\pm$0.016 & 0.630$\pm$0.023 \\
$^{78}$Kr/$^{84}$Kr 			& 1.06$\pm$0.14 & 1.57$\pm$0.24 & 1.317$\pm$0.099 & 1.88$\pm$0.11 & 1.48$\pm$0.32 \\
$^{80}$Kr/$^{84}$Kr 			& 8.52$\pm$0.52 & 17.77$\pm$0.82 & 13.99$\pm$0.54 & 12.39$\pm$0.60 & 18.8$\pm$1.0 \\
$^{82}$Kr/$^{84}$Kr 			& 25.2$\pm$1.4 & 31.6$\pm$1.7 & 26.7$\pm$1.1 & 26.2$\pm$1.0 & 29.1$\pm$1.3 \\
$^{83}$Kr/$^{84}$Kr 			& 25.37$\pm$0.99 & 28.4$\pm$1.4 & 25.82$\pm$0.69 & 26.6$\pm$1.1 & 26.6$\pm$1.6 \\
$^{86}$Kr/$^{84}$Kr 			& 29.4$\pm$1.3 & 28.1$\pm$2.0 & 29.82$\pm$0.86 & 29.7$\pm$1.5 & 29.8$\pm$1.9 \\
($^{80}$Kr/$^{82}$Kr)$_\text{excess}$ & 1.1$\pm$1.8 & 1.53$\pm$0.87 & 2.7$\pm$1.8 & 3.4$\pm$5.3 & 2.6$\pm$1.7 \\
\midrule
$^{132}$Xe 						& 1.3600$\pm$0.0098 & 0.7225$\pm$0.0094 & 0.773$\pm$0.010 & 0.865$\pm$0.013 & 0.862$\pm$0.014 \\
$^{124}$Xe/$^{132}$Xe 			& 0.538$\pm$0.035 & 0.642$\pm$0.060 & 0.688$\pm$0.054 & 0.636$\pm$0.053 & 0.608$\pm$0.055 \\
$^{126}$Xe/$^{132}$Xe 			& 0.607$\pm$0.035 & 0.88$\pm$0.10 & 0.876$\pm$0.056 & 0.672$\pm$0.070 & 0.718$\pm$0.042 \\
$^{128}$Xe/$^{132}$Xe 			& 8.62$\pm$0.16 & 8.79$\pm$0.22 & 8.86$\pm$0.46 & 8.52$\pm$0.29 & 8.78$\pm$0.30 \\
$^{129}$Xe/$^{132}$Xe 			& 112.1$\pm$1.1 & 111.2$\pm$1.6 & 116.2$\pm$2.4 & 111.8$\pm$2.0 & 113.2$\pm$2.5 \\
$^{130}$Xe/$^{132}$Xe 			& 16.27$\pm$0.18 & 16.53$\pm$0.55 & 16.99$\pm$0.36 & 16.13$\pm$0.33 & 16.48$\pm$0.48 \\
$^{131}$Xe/$^{132}$Xe 			& 81.72$\pm$0.74 & 83.4$\pm$1.6 & 84.5$\pm$1.9 & 80.9$\pm$1.3 & 82.8$\pm$2.2 \\
$^{134}$Xe/$^{132}$Xe 			& 38.99$\pm$0.49 & 39.5$\pm$1.0 & 40.19$\pm$0.95 & 38.85$\pm$0.82 & 38.30$\pm$0.84 \\
$^{136}$Xe/$^{132}$Xe 			& 33.06$\pm$0.39 & 33.58$\pm$0.96 & 34.06$\pm$0.80 & 32.21$\pm$0.73 & 31.2$\pm$1.1 \\
\botrule
 \end{tabular}}%
\footnotetext{$^{84}$Kr = 100, $^{132}$Xe=100 used for the ratios}
\end{table}

\begin{table}[ht]
\caption{Cosmogenic and radiogenic isotope concentrations, K, U, Th concentrations, shielding conditions (preatmospheric meteoroid radius and sample depth within the meteoroid), production rates P$_x$, cosmic ray exposure ages T$_x$, and U/Th-He and K-Ar gas retention ages T$_x$. $T_{26}$ ages were determined from the $^{26}$Al-$^{21}$Ne isotope pair method (cf. Methods \ref{sec:methods_gamma_spec}).}
\label{tab:CRE}%
\resizebox{\textwidth}{!}{\begin{tabular}{@{}llllll@{}}
\toprule
Sample & $^{3}$He$_{cos}$	& $^{21}$Ne$_{cos}$	& $^{38}$Ar$_{cos}$	& Radius\footnotemark[1] & Depth	\\
& 10$^{-8}$cm$^{3}$ STP/g & 10$^{-8}$cm$^{3}$ STP/g & 10$^{-8}$ cm$^{3}$ STP/g & cm & cm \\
\midrule
SPLV \#4  &	53.86 $\pm$ 0.15 & 12.860 $\pm$ 0.061 &	1.493 $\pm$ 0.027 & 35-50 & 21-40\\
SPLV \#26 &	51.04 $\pm$ 0.16 & 12.739 $\pm$ 0.065 &	1.568 $\pm$ 0.027 & 35-50 & 21-40 \\
SPLV \#13 &	50.59 $\pm$ 0.17 & 12.876 $\pm$ 0.079 &	1.461 $\pm$ 0.031 & 30-50 & 15-35 \\
SPLV \#10 & 53.28 $\pm$ 0.18 & 13.216 $\pm$ 0.072 & 1.689 $\pm$ 0.028 & 35-50 & 21-40 \\
SPLV \#1 & 51.66 $\pm$ 0.18 & 11.977 $\pm$ 0.078 & 1.271 $\pm$ 0.022	& 30-50	& 16-37	\\ 
\botrule
\end{tabular}}

\resizebox{\textwidth}{!}{\begin{tabular}{@{}llllllll@{}}
\toprule
Sample 	& P$_3$	& P$_{21}$ & 	P$_{38}$	& T$_3$	& T$_{21}$	& T$_{38}$ & T$_{26}$\\
 & \footnotesize{10$^{-8}$cm$^{3}$/(g*Myr)} & \footnotesize{10$^{-8}$cm$^{3}$/(g*Myr)} & \footnotesize{10$^{-8}$cm$^{3}$/(g*Myr)} & Myr & Myr & Myr & Myr\\
\midrule
SPLV \#4  &	1.93-2.03 &	0.42-0.44 &	0.045-0.046 & 26-28 &	29-31 &	32-34 & 29 $\pm$ 6\\
SPLV \#26 &	1.93-2.03 &	0.42-0.44 &	0.045-0.046	& 25-26 &	29-30 &	33-36 & 30 $\pm$ 6 \\
SPLV \#13 &	1.91-2.03 &	0.40-0.44 &	0.043-0.046 & 25-27 &	29-32 &	31-35 & 30 $\pm$ 6 \\
SPLV \#10 & 1.93-2.03 &	0.42-0.44 &	0.045-0.046 & 26-28 &	30-32 &	36-39 & 32 $\pm$ 9\\
SPLV \#1 & 1.91-2.03 &	0.41-0.44 &	0.043-0.046 & 25-27 &	27-30 &	27-30 & 30 $\pm$ 6 \\
\botrule
\end{tabular}}\\

\begin{tabular}{@{}llllllll@{}}
\toprule
Sample	& $^{4}$He$_{rad}$	& $^{40}$Ar$_{rad}$	& K\footnotemark[2]	& U\footnotemark[2]	& Th\footnotemark[2]	& T$_4$	& T$_{40}$ \\
& 10$^{-8}$cm$^{3}$ STP/g & 10$^{-8}$cm$^{3}$ STP/g  & ppm &  ppb & ppb & Gyr & Gyr \\
\midrule
SPLV \#4	& 573 $\pm$ 24	& 6370 $\pm$ 140	& 		& 	    &       & 1.8-1.9	& 4.4-4.5 \\
SPLV \#26	& 868 $\pm$ 23	& 6330 $\pm$ 140	& 		& 	    &       & 2.5-2.6	& 4.4-4.5 \\
SPLV \#13	& 965 $\pm$ 23	& 6400 $\pm$ 260	& 830	& 13	& 43	& 2.7-2.8	& 4.4-4.6 \\
SPLV \#10	& 927 $\pm$ 24	& 6050 $\pm$ 190	& 		& 	    &       & 2.6-2.8	& 4.3-4.4 \\
SPLV \#1	& 808 $\pm$ 23	& 6130 $\pm$ 250	& 		& 	    &       & 2.4-2.5	& 4.3-4.5 \\
\botrule
\end{tabular}

\footnotetext[1]{Upper limit set according to fireball observations}
\footnotetext[2]{Bulk chemical composition determined for SPLV \#1 via ICP-AES/ICP-MS, see Table \ref{tab:bulk_composition}, used for all samples here.}
\end{table}

\newpage
\clearpage
\color{black}

\newpage
\clearpage
\color{black}


\newpage

\clearpage
\newpage
\color{black}
\pagestyle{empty}    

\section*{Supplementary information}
\addcontentsline{toc}{section}{Supplementary information}

\begin{table}[h]
\caption{SPLV find coordinates of meteorites with known mass.}
\label{table:frag_coordinates}
\resizebox{0.99\textwidth}{!}{\begin{tabular}{@{}llllll@{}}
\toprule
  Number & Discoverer(s) & Mass (g) & Latitude  ($\degree$N) & Longitude  ($\degree$E) & Date found (2023)\\ 
\midrule 
1  & Thierry Monter & 490 & 49.796198 & 0.859827 & Feb. 26 \\
2  & Steve Arnold & 175.2 & 49.808110 & 0.827920 & Feb. 19 \\
3  & Łukasz \& Magdalena Smuła & 148.20 & 49.805733 & 0.838517 & \\
4  & Loïs Leblanc-Rappe & 92.6 & 49.820911 & 0.826400 & Feb. 15*\\
5  & Bruno Freire Barroca & 28 & 49.826388 & 0.804722 &  Feb. 20 \\
6  & Juliette Loubière & 23.8 & 49.822768 & 0.799881 & Feb. 18 \\
7  & Renaud Trangosi & 23.6 & 49.825577 & 0.797596 & Feb. 18 \\
8  & Vincent Jacques & 22.18 & 49.026883 & 0.796936 & Feb. 24 \\
9  & Bruno Freire Barroca & 18 & 49.825000 & 0.798056 & Feb. 18 \\
10 & Pierre Beck & 17 & 49.823086 & 0.798431 & Feb. 19 \\
11 & Łukasz \& Magdalena Smuła  & 13.99 & 49.819433 & 0.799517 & \\
12 & Łukasz \& Magdalena Smuła  & 13.43 & 49.826017 & 0.794583 & \\
13 & Iliès et Amine & 11 & 49.825775 & 0.795238 & Feb. 18 \\
14 & Arnaud Leroy & 9 & 49.832984 & 0.786048 & Feb. 17 \\
15 & Unknown & 8.7 & 49.841830 & 0.780835 & \\
16 & Łukasz \& Magdalena Smuła  & 8.56 & 49.826467 & 0.785250 & \\
17 & Pierre Vernazza & 7.6 & 49.831460 & 0.786022 & Feb. 17 \\
18 & Pierre Vernazza & 7.2 & 49.832833 & 0.783355 & Feb. 17 \\
19 & Łukasz \& Magdalena Smuła  & 7.07 & 49.825300 & 0.795883 & \\
20 & Zsolt Kereszty & 6.6 & 49.846667 & 0.787500 & Mar.  1 \\
21 & Bruno Freire Barroca & 6 & 49.826666 & 0.788333 & Feb. 18 \\
22 & Tom Hughes & 5.89 & 49.834467 & 0.787333 & Feb. 24 \\
23 & Unknown & 5.80 & 49.842799 & 0.783038 & \\
24 & Łukasz \& Magdalena Smuła & 5.53 & 49.841033 & 0.781800 & \\
25 & Aidar Zakirov & 5.40 & 49.831010 & 0.786250 & Feb. 19 \\
26 & Ludovic Ferrière & 5.12 & 49.796198 & 0.859827 & Feb. 26 \\
27 & Unknown & 5.10 & 49.842883 & 0.781866 & \\
28 & Bruno Freire Barroca & 5 & 49.833333 & 0.783056 & Feb. 20 \\
29 & Łukasz \& Magdalena Smuła  & 4.88 & 49.840100 & 0.785167 & \\
30 & Lucie Maquet & 4.00 & 49.832306 & 0.784500 & Feb. 17 \\
31 & Łukasz \& Magdalena Smuła  & 3.96 & 49.837067 & 0.778000 & \\
32 & Peter Jenniskens & 3.50 & 49.832683 & 0.781639 & Feb. 16 \\
33 & Unknown & 3.30 & 49.842799 & 0.783038 & \\
34 & Łukasz \& Magdalena Smuła  & 3.07 & 49.835850 & 0.776767 & \\
35 & Łukasz \& Magdalena Smuła  & 2.94 & 49.837883 & 0.777483 & \\
36 & Łukasz \& Magdalena Smuła  & 2.94 & 49.837250 & 0.776083 & \\
37 & Łukasz \& Magdalena Smuła  & 2.58 & 49.838783 & 0.778583 & \\
38 & Unknown & 2.40 & 49.840021 & 0.780323 & \\
39 & Unknown & 2.40 & 49.842757 & 0.779150 & \\
40 & Łukasz \& Magdalena Smuła  & 2.35 & 49.839283 & 0.787667 & \\
41 & Łukasz \& Magdalena Smuła  & 2.34 & 49.840317 & 0.778750 & \\
42 & Łukasz \& Magdalena Smuła  & 2.15 & 49.836867 & 0.779717 & \\
43 & François Colas & 2.00 & 49.824075 & 0.800605 & Feb. 17 \\
44 & Łukasz \& Magdalena Smuła  & 1.87 & 49.836683 & 0.776033 & \\
45 & Łukasz \& Magdalena Smuła  & 1.81 & 49.838633 & 0.774633 & \\
46 & Łukasz \& Magdalena Smuła  & 1.63 & 49.836844 & 0.776967 & \\
47 & Vincent Jacques & 1.51 & 49.796198 & 0.859827 & Feb. 28 \\
48 & Łukasz \& Magdalena Smuła  & 1.09 & 49.836717 & 0.774950 & \\
49 & Łukasz \& Magdalena Smuła  & 0.84 & 49.838483 & 0.776217 & \\
  \botrule
\end{tabular}}
\footnotetext{ *) First find.}
\end{table}

\begin{table}[ht]
\caption{SPLV find coordinates of meteorites (continued)$^*$. }\label{table:frag_coordinates2}
\resizebox{0.99\textwidth}{!}{\begin{tabular}{@{}llllll@{}}
\toprule
  Number & Discoverer(s) & Mass (g) & Latitude  ($\degree$N) & Longitude  ($\degree$E) & Date found (2023)\\ 
\midrule 
50 &  Filip Nikodem, Andrzej Owczarzak, Michal Nebelski & 11.0 & 49.834639 &  0.785881  & \\
51 &  Filip Nikodem, Andrzej Owczarzak, Michal Nebelski & 8.4 & 49.834447  &  0.786566  & \\
52 &  Filip Nikodem, Andrzej Owczarzak, Michal Nebelski & 7.8 & 49.841830  &  0.780835  & \\
53 &  Filip Nikodem, Andrzej Owczarzak, Michal Nebelski & 7.6 & 49.843497  &  0.781363  & \\
54 &  Filip Nikodem, Andrzej Owczarzak, Michal Nebelski & 7.1 & 49.833483  &  0.786520  & \\
55 &  Filip Nikodem, Andrzej Owczarzak, Michal Nebelski & 7.1 & 49.843808  &  0.782136  & \\
56 &  Filip Nikodem, Andrzej Owczarzak, Michal Nebelski & 7.0 & 49.841300  &  0.781199  & \\
57 &  Filip Nikodem, Andrzej Owczarzak, Michal Nebelski & 6.4 & 49.838949  &  0.786212  & \\
58 &  Filip Nikodem, Andrzej Owczarzak, Michal Nebelski & 6.0 & 49.833380  &  0.785739  & Feb. 23\\
59 &  Filip Nikodem, Andrzej Owczarzak, Michal Nebelski & 6.0 & 49.840947  &  0.786388  & \\
60 &  Filip Nikodem, Andrzej Owczarzak, Michal Nebelski & 5.5 & 49.834507  &  0.777901  & \\
61 &  Filip Nikodem, Andrzej Owczarzak, Michal Nebelski & 5.1 & 49.834992  &  0.776799  & Mar. 22\\
62 &  Filip Nikodem, Andrzej Owczarzak, Michal Nebelski & 4.8 & 49.841576  &  0.781837  & \\
63 &  Filip Nikodem, Andrzej Owczarzak, Michal Nebelski & 4.6 & 49.834754  &  0.775359  & \\
64 &  Filip Nikodem, Andrzej Owczarzak, Michal Nebelski & 4.6 & 49.840684  &  0.785033  & \\
65 &  Filip Nikodem, Andrzej Owczarzak, Michal Nebelski & 4.4 & 49.835178  &  0.776030  & \\
66 &  Filip Nikodem, Andrzej Owczarzak, Michal Nebelski & 4.4 & 49.834886  &  0.778164  & \\
67 &  Filip Nikodem, Andrzej Owczarzak, Michal Nebelski & 4.2 & 49.831296  &  0.780495  & \\
68 &  Filip Nikodem, Andrzej Owczarzak, Michal Nebelski & 4.2 & 49.841019  &  0.785514  & \\
69 &  Filip Nikodem, Andrzej Owczarzak, Michal Nebelski & 4.0 & 49.843167  &  0.782031  & \\
70 &  Filip Nikodem, Andrzej Owczarzak, Michal Nebelski & 3.9 & 49.835153  &  0.776012  & \\
71 &  Filip Nikodem, Andrzej Owczarzak, Michal Nebelski & 3.8 & 49.842309  &  0.784584  & \\
72 &  Filip Nikodem, Andrzej Owczarzak, Michal Nebelski & 3.6 & 49.840684  &  0.785033  & \\
73 &  Filip Nikodem, Andrzej Owczarzak, Michal Nebelski & 3.6 & 49.830990  &  0.780519  & \\
74 &  Filip Nikodem, Andrzej Owczarzak, Michal Nebelski & 3.4 & 49.836082  &  0.775542  & \\
75 &  Filip Nikodem, Andrzej Owczarzak, Michal Nebelski & 3.3 & 49.835656  &  0.777676  & \\
76 &  Filip Nikodem, Andrzej Owczarzak, Michal Nebelski & 3.2 & 49.840800  &  0.779000  & \\
77 &  Filip Nikodem, Andrzej Owczarzak, Michal Nebelski & 2.8 & 49.842345  &  0.783284  & \\
78 &  Filip Nikodem, Andrzej Owczarzak, Michal Nebelski & 2.6 & 49.839598  &  0.773529  & \\
79 &  Filip Nikodem, Andrzej Owczarzak, Michal Nebelski & 2.6 & 49.835581  &  0.778591  & \\
80 &  Filip Nikodem, Andrzej Owczarzak, Michal Nebelski & 2.5 & 49.839781  &  0.773869  & \\
81 &  Filip Nikodem, Andrzej Owczarzak, Michal Nebelski & 2.3 & 49.845230  &  0.785517  & \\
82 &  Filip Nikodem, Andrzej Owczarzak, Michal Nebelski & 2.3 & 49.840695  &  0.779795  & \\
83 &  Filip Nikodem, Andrzej Owczarzak, Michal Nebelski & 2.2 & 49.839777  &  0.774157  & \\
84 &  Filip Nikodem, Andrzej Owczarzak, Michal Nebelski & 2.2 & 49.840551  &  0.772411  & \\
85 &  Filip Nikodem, Andrzej Owczarzak, Michal Nebelski & 2.1 & 49.842757  &  0.779150  & \\
86 &  Filip Nikodem, Andrzej Owczarzak, Michal Nebelski & 2.1 & 49.843978  &  0.784420  & \\
87 &  Filip Nikodem, Andrzej Owczarzak, Michal Nebelski & 2.1 & 49.835252  &  0.778973  & \\
88 &  Filip Nikodem, Andrzej Owczarzak, Michal Nebelski & 2.0 & 49.840033  &  0.772760  & \\
89 &  Filip Nikodem, Andrzej Owczarzak, Michal Nebelski & 1.9 & 49.841805  &  0.784344  & \\
90 &  Filip Nikodem, Andrzej Owczarzak, Michal Nebelski & 1.8 & 49.842062  &  0.779259  & \\
91 &  Filip Nikodem, Andrzej Owczarzak, Michal Nebelski & 1.7 & 49.841382  &  0.783850  & \\
92 &  Filip Nikodem, Andrzej Owczarzak, Michal Nebelski & 1.7 & 49.838961  &  0.784503  & \\
93 &  Filip Nikodem, Andrzej Owczarzak, Michal Nebelski & 1.4 & 49.842345  &  0.783284  & \\
94 &  Filip Nikodem, Andrzej Owczarzak, Michal Nebelski & -.- & 49.842883  &  0.781866  & \\
95 &  Filip Nikodem, Andrzej Owczarzak, Michal Nebelski & -.- & 49.842799  &  0.783038  & \\
96 &  Filip Nikodem, Andrzej Owczarzak, Michal Nebelski & -.- & 49.834913  &  0.787170  & \\
97 &  Filip Nikodem, Andrzej Owczarzak, Michal Nebelski & -.- & 49.833806  &  0.786639  & \\
98 &  Filip Nikodem, Andrzej Owczarzak, Michal Nebelski & -.- & 49.834306  &  0.786556  & \\
99 & unknown & 8 & 49.828257 & 0.789998 & Feb. 25\\
100 & unknown & 20 & 49.826278 & 0.797283 & Feb. 28\\
  \botrule
\end{tabular}}
\footnotetext{ *) Data provided by F. Nikodem.}
\end{table}

\begin{table}[ht]
\caption{Bulk rock elemental compositions of SPLV \#1, measured following the procedure described in \cite{Barrat2012}. A 250 mg aliquot was measured by ICP-AES (Sample 1) and a 120 mg aliquot analyzed by ICP-MS (Sample 2).}
\label{tab:bulk_composition}
\centering
\begin{tabular}{@{}lr|lrlr@{}}
\toprule
Sample	&	1	&		&		&	2	  &		\\
        & wt\%	&		&	 	&	µg/g  &	\\
\midrule
SiO2	&	38.96	&	Li	&	1.98	&	La	&	0.344	\\
TiO2	&	0.09	&	Be	&	0.0293	&	Ce	&	0.898	\\
Al2O3	&	1.70	&	Al	&	11.9	&	Pr	&	0.134	\\
FeO	&	27.24	&	P	&	1.20	&	Nd	&	0.672	\\
MnO	&	0.33	&	K	&	983	&	Sm	&	0.219	\\
MgO	&	26.27	&	Ca	&	13.4	&	Eu	&	0.0823	\\
CaO	&	1.66	&	Sc	&	8.91	&	Gd	&	0.296	\\
Na2O	&	0.88	&	Ti	&	605	&	Tb	&	0.0543	\\
K2O	&	0.10	&	V	&	68.7	&	Dy	&	0.370	\\
P2O5	&	0.30	&	Mn	&	3086	&	Ho	&	0.0820	\\[-0.1cm]
\cmidrule{1-2} & \\[-0.5cm]
\textbf{Total}	&	97.54	&	Co	&	532	&	Er	&	0.242	\\
	&	µg/g &	Cu	&	70.5	&	Tm	&	0.0369	\\[-0.1cm]
\cmidrule{1-2} & \\[-0.5cm]
Cr	&	1555	&	Zn	&	53.6	&	Yb	&	0.240	\\
Ni	&	11348	&	Ga	&	4.86	&	Lu	&	0.0361	\\
	&		&	Rb	&	3.26	&	Hf	&	0.156	\\
	&		&	Sr	&	11.2	&	Ta	&	0.0190	\\
	&		&	Y	&	2.34	&	W	&	0.118	\\
	&		&	Zr	&	5.04	&	Pb	&	0.0195	\\
	&		&	Nb	&	0.484	&	Th	&	0.0431	\\
	&		&	Cs	&	0.0277	&	U	&	0.0126	\\
	&		&	Ba	&	3.54	&		&		\\
 \cmidrule{3-6}
\end{tabular}
\end{table}

\begin{table}[ht]
  \caption{Massic activities of cosmogenic radionuclides (in dpm~kg$^{-1}$, uncertainties at 1 sigma) measured in SPLV fragments by non-destructive gamma-ray spectrometry (GS), corrected back to the date of the meteorite fall on February 13, 2023. } 
    \label{tab:radionuclide_activities}
    \centering
    \begin{tabular}{llllll}
\toprule
 GS  & Half-life & \#2 & \#4 & \#10 \\
\midrule
$^{22}$Na & 2.6022 yr & 112 $\pm$ 6 & 114 $\pm$ 8 & 115 $\pm$ 13 \\
$^{26}$Al & 0.717 Myr & 57.6 $\pm$ 5.4 & 60.2 $\pm$ 5.9 & 61.3 $\pm$ 7.8 \\
$^{46}$Sc & 83.788 d &  & 8.1 $\pm$ 1.0 & & \\ 
$^{48}$V & 16 d &  & 18.0 $\pm$ 2.2 & & \\
$^{54}$Mn & 312.13 d& 124 $\pm$ 8 & 129 $\pm$ 8 & 152 $\pm$ 16 \\
$^{57}$Co & 271.8 d & 13.8 $\pm$ 3.4 & 15.0 $\pm$ 3.7 &  & \\
$^{58}$Co & 70.83 d &  10.2 $\pm$ 2.2 & 10.8 $\pm$ 2.2 &  & \\
$^{60}$Co & 5.2711 yr & 7.3 $\pm$ 1.7 & 10.1 $\pm$ 1.6 & 12.4 $\pm$ 3.8 \\ 
$^{22}$Na/$^{26}$Al & & 1.94 $\pm$ 0.17 & 1.89 $\pm$ 0.21 & 1.88 $\pm$ 0.32\\
\botrule
    \end{tabular}
\end{table}

\begin{table}[ht]
  \caption{Ratios and resulting massic radionuclide activities measured by IAMS in five SPLV samples (in dpm/kg). Total uncertainties (at 1-$\sigma$) include also those from the standards and 5\% for stable Al (1.19\%) and Ca (1.19\%) determination for massic activities. } 
    \label{tab:IAMS}
 \resizebox{0.99\textwidth}{!}{   \begin{tabular}{lllllll}
    \toprule
    sample & $^{26}$Al/$^{27}$Al ($\times 10^{-10}$) & $^{26}$Al & $^{41}$Ca/$^{40}$Ca ($\times 10^{-12}$) & $^{41}$Ca & Radius (cm) & Depth (cm) \\
   \midrule
    \#1 & 1.169 $\pm$ 0.047 & 58.1 $\pm$ 3.7 & 2.57 $\pm$ 0.36  &   5.65 $\pm$ 0.84 & $\ge$20 & 8-30 \\
    \#4 &  1.041 $\pm$ 0.032 & 51.7 $\pm$ 3.0 &  2.69 $\pm$ 0.36  &   5.92 $\pm$ 0.85 & $\ge$12.5 & 2.5-15  \\
    \#10  & 1.039 $\pm$ 0.041 & 51.6 $\pm$ 3.3 &  4.01 $\pm$ 0.86  &   8.8 $\pm$ 1.9 & $\ge$12.5 & 2.5-15   \\ 
    \#13 &  1.126 $\pm$ 0.042 & 55.9 $\pm$ 3.5 &  3.97 $\pm$ 0.51  &   8.7 $\pm$ 1.2 & $\ge$15 & 4-20  \\ 
    \#26 &   1.053 $\pm$ 0.039 & 52.3 $\pm$ 3.3 &  2.94 $\pm$ 0.64  &   6.5 $\pm$ 1.5 & $\ge$12.5 & 2.5-15  \\
   \botrule
    \end{tabular}}
\end{table}

\begin{table}[ht]
\caption{Bulk rock density of SPLV fragments 2, 4, 7 \& 30, based on photogrammetry and laser scanning 3D reconstruction. The average bulk rock density based on laser scans is 3.2940 g/cm$^{3}$ with a standard deviation of 0.0020 g/cm$^{3}$. The average bulk rock density based on the photogrammetry models give a density of 3.3529 g/cm$^{3}$ with a standard deviation of 0.080 g/cm$^{3}$.}\label{tab:density}%
    \centering
\begin{tabular}{@{}lllll@{}}
\toprule
 Fragment & Method & Mass (g) & Volume (cm$^{3}$) & Density (g/cm$^{3}$)  \\
\midrule
2 & Photogrammetry & 175.0 & 53.6565 & 3.2615   \\
\midrule
4 & Photogrammetry & 93.9 & 27.188 & 3.4537   \\
 & Laser scan & & 28.4977 & 3.2950 \\
\midrule
6 & Photogrammetry & 23.628 & 7.0200 & 3.3658   \\
 & Laser scan & & 7.1781 & 3.2917 \\
\midrule
30 & Photogrammetry & 3.958 & 1.1884 & 3.3306   \\
 & Laser scan & & 1.2011 & 3.2969 \\
\botrule
\end{tabular}
\end{table}

\begin{table}[ht]
\centering
\caption{Probabilities (in \%) of originating from specific sources, calculated using METEOMOD (\url{https://sirrah.troja.mff.cuni.cz/~mira/meteomod/meteomod.html}) for the orbits of Saint-Pierre-le-Viger (SPLV, $a = 1.629\,{\rm au}$, $i = 3.418^\circ$) and Novo Mesto (NM, $a = 1.451\,{\rm au}$, $i = 8.755^\circ$). The top table includes all asteroid families with spectral types compatible with H, L, and LL chondrites, as identified in \cite{Broz2024s}. The lower table restricts possible sources to those consistent with L chondrite-like compositions. Both analyses include the recently identified Massalia$_2$ and Koronis$_2$ subpopulations \cite{Broz2024s,Marsset2024s}. When considering L chondrite-like sources only, the highest probability for both SPLV and NM corresponds to Massalia$_2$, a subpopulation likely formed by a cratering or reaccumulation event on asteroid (20) Massalia approximately 40 million years ago \cite{Marsset2024s}.\color{black}} \label{tab:meteomod}
\begin{tabular}{rlrr}
\# & Family (Taxonomy) & SPLV & NM  \\
\hline
{\bf 20'} & {\bf Massalia$_2$ (L)} & {\bf 39.90} & {\bf 51.66} \\
158' & Koronis$_2$ (H)         & 36.06 & 23.11 \\
 832 & Karin (H)            & 10.62 &  6.81 \\
   8 & Flora (LL)           &  8.84 & 12.89 \\
  20 & Massalia (L)         &  1.87 &  2.43 \\
 135 & Hertha (LL)          &  1.77 &  2.01 \\
   3 & Juno (L/LL)          &  0.65 &  0.00 \\
 158 & Koronis (H)          &  0.28 &  0.34 \\
 847 & Agnia (H)            &  0.00 &  0.71 \\
  25 & Phocaea (H)          &  0.00 &  0.03 \\
1272 & Gefion (L)           &  0.00 &  0.00 \\
  15 & Eunomia (LL)         &  0.00 &  0.00 \\
 170 & Maria (H)            &  0.00 &  0.00 \\
 808 & Merxia (H)           &  0.00 &  0.00 \\
\hline   
\hline
\end{tabular}
\footnotetext{}
\vspace{0.2cm}
\begin{tabular}{rlrr}
\# & Family (Taxonomy) & SPLV & NM  \\
\hline
  20 & Massalia (L)         &  4.42 &  4.49 \\
{\bf 20'} & {\bf Massalia$_2$ (L)} & {\bf 94.04} & {\bf 95.51} \\
   3 & Juno (L/LL)          &  1.54 &  0.00 \\
1272 & Gefion (L)           &  0.00 &  0.00 \\
\hline   
\hline
\end{tabular}
\end{table}


\newpage
\clearpage

\newpage

\end{document}